\documentclass[twocolumn]{aastex6}
\usepackage{amsmath}
\usepackage{amssymb}
\usepackage{amsthm}
\usepackage{natbib,aasdefs,url,bm}
\usepackage{array}
\usepackage{float}
\usepackage{graphicx}
\usepackage{color}

\newcommand{\ccmajor}{}
\newcommand{\ccminor}{}

\newcounter{ichi}
\newcounter{ni}
\newcounter{san}
\newcounter{yon}
\def\be{\begin{equation}}
\def\ee{\end{equation}}
\def\ba{\begin{eqnarray}}
\def\ea{\end{eqnarray}}

\newcommand{\refsec}[1]{\S\ref{sec:#1}}

\def\km{}

\slugcomment{}

\shorttitle{Cumulative Neutrino and Gamma-Ray Backgrounds from Halo and Galaxy Mergers}
\shortauthors{Yuan et al.}

\begin{document}

\title{Cumulative Neutrino and Gamma-Ray Backgrounds from Halo and Galaxy Mergers}
\author{Chengchao Yuan\altaffilmark{1}}\email{cxy52@psu.edu}
\author{Peter M\'esz\'aros\altaffilmark{1}}
\author{Kohta Murase\altaffilmark{1,2} }
\author{Donghui Jeong\altaffilmark{1} }
\altaffiltext{1}{Department of Physics; Department of Astronomy and Astrophysics; Center for Particle and Gravitational Astrophysics, The Pennsylvania State University, University Park, PA 16802, USA}
\altaffiltext{2}{Center for Gravitational Physics, Yukawa Institute for Theoretical Physics, Kyoto University, Kyoto 606-8502, Japan}

%\email{cxy52@psu.edu}

\begin{abstract}
The merger of dark matter halos and the gaseous structures embedded in them, such as proto-galaxies, galaxies, and groups and clusters of galaxies, results in strong shocks that are capable of accelerating cosmic rays (CRs) to {\ccmajor $\gtrsim10~\rm PeV$}. These shocks will produce high-energy neutrinos and $\gamma$-rays through inelastic $pp$ collisions. In this work, we study the contributions of these halo mergers to the diffuse neutrino flux and to the non-blazar portion of the extragalactic $\gamma$-ray background. We formulate the redshift dependence of the shock velocity, galactic radius, halo gas content and galactic/intergalactic magnetic fields over the dark matter halo distribution up to a redshift $z=10$. We find that high-redshift mergers contribute a significant amount of the cosmic-ray luminosity density, and the resulting neutrino spectra could explain a large part of the observed diffuse neutrino flux {\ccmajor above 0.1 PeV up to $\sim$ PeV}. We also show that our model can somewhat alleviate tensions with the extragalactic $\gamma$-ray background. First, since a larger fraction of the {\ccminor CR luminosity density} comes from high redshifts, the accompanying $\gamma$-rays are more strongly suppressed through $\gamma\gamma$ annihilations with the cosmic microwave background (CMB) and the extragalactic background light (EBL). Second, mildly radiative-cooled shocks may lead to a harder CR spectrum with spectral indices of $1.5\lesssim s\lesssim2.0$. Our study suggests that halo mergers, a fraction of which may also induce starbursts in the merged galaxies, can be promising neutrino emitters without violating the existing {\it Fermi} $\gamma$-ray constraints on the non-blazar component of the extragalactic $\gamma$-ray background.
\end{abstract}

%\keywords{gamma-ray burst: general --- gravitational waves --- radio continuum: general --- stars: magnetars --- stars: neutron}

%%%%%%%%%%%%%%%%%%%%%%%%%%%%%%%%%%%%%%%%%%%%%%%%%%%%%%%%%%%%%%%%%%%%%%%%%%%%%
\section{Introduction}
\label{sec:into}
%%%%%%%%%%%%%%%%%%%%%%%%%%%%%%%%%%%%%%%%%%%%%%%%%%%%%%%%%%%%%%%%%%%%%%%%%%%%%

Neutrino astrophysics has made substantial progress since the {\ccminor IceCube
Neutrino Observatory in Antarctic} \citep[e.g.,][for reviews]{doi:10.1146/annurev-nucl-102313-025321,Halzen:2016gng} (hereafter, IceCube) 
was completed. During the last half decade, scores of high-energy (HE) astrophysical
neutrinos with energies between $\sim10\ \rm TeV$ and a few $\rm PeV$ have 
been detected by IceCube, and the number keeps {\ccminor growing} 
\citep{Aartsen:2013bka,Aartsen:2013jdh,Aartsen:2014gkd,aartsen2015combined}. 
The arrival directions of these neutrinos are compatible with an isotropic 
distribution even in the $10-100$ TeV range, suggesting that 
a large part of these diffuse neutrinos come from extragalactic sources. 
%in the galactic coordinate
Non-observation of diffuse Galactic $\gamma$-rays from the Galactic plane and other extended regions independently suggest that the 
Galactic contribution (e.g., by Fermi bubbles or local supernova remnants) is unlikely to be 
dominant \citep{Ahlers:2013xia,Apel:2017ocm,Abeysekara:2017wzt,Abeysekara:2017jxs}. 
However, despite extensive efforts, the physical nature of the sources of the diffuse neutrinos still remains in dispute. 
Possible candidates include gamma-ray bursts (GRBs) \citep[e.g.,][]{waxman1997high,meszaros2001tev,murase2008prompt,wang2009prompt,baerwald2013uhecr,Bustamante:2014oka,tamborra2016inspecting}, 
low-power GRBs \citep{murase2006high,gupta2007neutrino,murase2013tev,xiao2014neutrino,xiao2015tev,senno2016choked,Denton:2017jwk}, radio-loud active galactic nuclei (AGNs) \citep[e.g.,][]{Mannheim:1995mm,Halzen:1997hw,anchordoqui2008high,murase2014diffuse,dermer2014photopion,Tjus:2014dna,petropoulou2015photohadronic,Padovani:2015mba,Blanco:2017bgl}, radio-quiet/low-luminosity AGNs \citep{stecker1991high,alvarez2004high,stecker2013pev,Kimura:2014jba}, and AGNs embedded in galaxy clusters and groups\footnote{Groups of galaxies are smaller clusters, 
numbering from a few to dozens of galaxies.}.  It is generally accepted that the bulk of astrophysical neutrinos are generated by charged pion ($\pi^\pm$) decays, and 
that these pions are the secondaries from cosmic ray (CR) particles undergoing hadronuclear ($pp$) or photohadronic ($p\gamma$) interactions 
between the CRs and ambient target gas nuclei or photons. 
Meanwhile, these collisions also lead unavoidably to neutral pions ($\pi^0$) as well, which 
subsequently decay into a pair of $\gamma$-rays. Hence, the diffuse neutrino flux is expected to have an intimate connection with the diffuse CR 
and $\gamma$-ray backgrounds, and multi-messenger analyses need to be applied to constrain the origin of these diffuse high-energy cosmic particle fluxes \citep{murase2013testing,Murase:2015xka,bechtol2017evidence}. 

Galaxy clusters and groups have been considered as promising candidate sources of IceCube's neutrinos, and CR accelerators can be 
not only AGNs but also intragalactic sources, accretion shocks, and mergers of clusters and groups ~\citep[e.g.,][]{murase2008cosmic,fang2018linking}.
Star-forming and starburst galaxies (SFGs \& SBGs, respectively) have also been suggested as promising 
candidates for HE neutrino sources \citep[e.g.,][]{loeb2006cumulative,murase2013testing, 
tamborra2014star,anchordoqui2014icecube,chang2014diffuse,chang2015star,senno2015extragalactic,Chakraborty:2015sta}. 
In particular, starburst galaxies have dense gaseous environments {\ccminor and} have been of interest as efficient CR reservoirs. 
Previous studies have assumed not only supernova and hypernova remnants (SNRs \& HNRs, respectively) but also galaxy mergers, 
disk-driven outflows and possible weak jets from AGNs as CR accelerators embedded in the star-forming galaxies
\citep{murase2013testing,tamborra2014star,kashiyama2014galaxy,senno2015extragalactic,Chakraborty:2015sta,Lamastra:2017iyo}. 
{Hypernovae (HNe) are a subclass of Type Ib/c supernovae (SNe), essentially a hyper-energetic 
version of Ib/c SNe. The typical ejecta energy of HNe is $~10^{52}$~erg, which is one order of magnitude larger than for SNe. 
Like SNRs, a hypernova remnant (HNR) leads to an extended structure that results from a hypernova explosion.}
In any case, an important constraint on such models is provided by the extragalactic $\gamma$-ray background (EGB) in the 
$100~{\rm MeV}-820~{\rm GeV}$ range, derived from the observation by the {\it Fermi}-LAT satellite \citep{ackermann2015spectrum}. 
Recent studies of the blazar flux distribution at $\gamma$-ray energies above 
$50~{\rm GeV}$ indicate that  {\ccminor{blazars account for}} $86^{+16}_{-14}\%$ of the 
total EGB flux \citep{Fermi16}. This provides a strong constraint, 
namely, only a fraction $\lesssim30\%$, with a best fit of $14\%$, can be ascribed 
to any remaining non-blazar component of the EGB \citep[see also][]{Lisanti:2016jub}. 

With this constraint, the SBG scenario is apparently disfavored as the dominant origin of IceCube neutrinos {\ccminor\citep{bechtol2017evidence}}. However, so far, this conclusion depends on the interpretation of the medium-energy neutrino data in the $10-100$ TeV range. For example, the cumulative neutrino background may consist of two components, in which the high-energy data above $\sim100$ TeV can be explained 
by the SBGs. On the other hand, the $10-100$ TeV component motivates CR accelerators that are ``dark'' in $\gamma$-rays \citep{Murase:2015xka} 
to satisfy multi-messenger constraints. The $\gamma$-rays may be attenuated inside their sources, or they might 
be absorbed during the propagation. Possible candidates include {\ccminor choked-jet GRBs} 
or high-redshift sources such as Pop-III HNRs embedded in starbursts \citep{Xiao+16nuhn}. 

{\ccmajor 
In this paper, we focus on halo mergers as an origin of HE neutrinos.  In the standard hierarchical galaxy 
formation scenario, galaxies form inside extended dark matter halos. When dark matter halos merge, 
the galaxies in these halos also merge, and the collision of the cold gas in the merging galaxies leads to 
shocks on a galactic scale in the galactic interstellar medium (ISM) gas. Later in the process of cosmological 
structure formation, at lower redshifts where galaxy groups and galaxy clusters have started forming, mergers among the dark 
matter halos containing these groups and clusters are also expected, These mergers are very energetic and 
result in shocks in the intergalactic medium (IGM) gas of the participating groups/clusters. One vivid example 
is the Bullet Cluster \citep{clowe2004weak,markevitch2004direct}}. Both these galactic and group/cluster shocks 
can accelerate CRs. The subsequent $pp$ collisions between the shock-accelerated CRs and the thermal atomic 
nuclei in the gaseous environment is the major mechanism that generates HE neutrinos in these systems.

Here, we consider this scenario {\ccmajor of both galactic scale shocks in the galactic ISM and group/cluster 
scale shocks in the intergalactic gas across redshifts. Whereas in a previous study \citep{kashiyama2014galaxy}
only major galaxy mergers (mergers of two galaxies of approximately the same size) at $z\sim 1$ were considered,
here take into account the redshift evolution of the halo merger rate, and consider both galaxy and cluster
mergers, including both major and minor mergers (the latter being those where the participating galaxies or clusters
have mass ratios $\zeta \neq 1$). We calculate the CR productions in the corresponding shocks at redshifts 
$0\leq z \leq 10$ and we find that high redshift ($z\gtrsim1-2$) halo mergers contribute a significant part of the
observed diffuse HE neutrinos and $\gamma$-rays. 
}
%We also show that such galactic halo mergers help the tension with the EGB, since the $\gamma$-rays generated at high redshifts 
%are sufficiently attenuated due to the enhanced $\gamma\gamma$ interactions. 
%In addition, because galaxies are statistically smaller at early times, the CR diffusion time is also shorter. As a result, 
%both the neutrino production efficiency and $\gamma$-ray input in the $\lesssim1~{\rm TeV}$-range are limited.
%This effect lowers the $\gamma$-ray intensity observed by $Fermi$, and, therefore, makes the early galactic halo mergers even darker source of HE neutrinos.
%In addition, considering the merger history 
%of galaxies and halos, early galaxies are smaller than present-day galaxies, 
%which results in limiting the neutrino production efficiency in the energy 
%range $\lessapprox\rm TeV$ by lowering the CR diffusion time. 
%This effect also lowers the gamma-ray input in this range observed by $Fermi$, 
%making the early galactic halo mergers even darker source of HE neutrinos.

This paper is organized as follows. In \S\ref{sec:halo}, we introduce 
the halo mass function and the halo distribution that is used in the following 
sections. The merger rate and the CR energy input rate are given in 
\S\ref{sec:CR}. In \S\ref{sec:nugam} we discuss the redshift dependence of the 
CR maximum energies and the neutrino {\ccminor product efficiency}, and we calculate the 
resulting {\ccminor neutrino and the $\gamma$-ray fluxes}. The results and implications 
are discussed in \S\ref{sec:dis}. 
Throughout, we assume a standard flat-$\Lambda$CDM universe with 
{\ccminor present-day density parameter} $\Omega_{m,0}=0.3$ and 
Hubble parameter $H_0=71.9\rm\ km\ s^{-1}\ Mpc^{-1}$\citep{bonvin2017h0licow}. 

%Modifications to the previous version: 

%In eqn (12), the energy in $\gamma\gamma$ optical depth should be $(1+z)\varepsilon_\gamma$ instead of $\varepsilon_\gamma$ since the optical depths given by J. Finke ($z<5$) and Y. Inoue ($5<z<10$) are functions of observed energies.

%In eqn (10), there is no $\Omega_{m,0}^{0.45}$

%%%%%%%%%%%%%%%%%%%%%%%%%%%%%%%%%%%%%%%%%%%%%%%%%%%%%%%%%%%%%%%%%%%%%%%%%%%%%
\section{Halo Mass Function}
\label{sec:halo}
%%%%%%%%%%%%%%%%%%%%%%%%%%%%%%%%%%%%%%%%%%%%%%%%%%%%%%%%%%%%%%%%%%%%%%%%%%%%
%%%%%%%%%%%%%%%%%%%%%%%%%%%%%%%%%%%%%%%%%%%%%%%%%%%%%%%%%%%%%%%%%%%%%%%5
\begin{figure}[t!]
{\centering
	\includegraphics[width=0.45\textwidth]{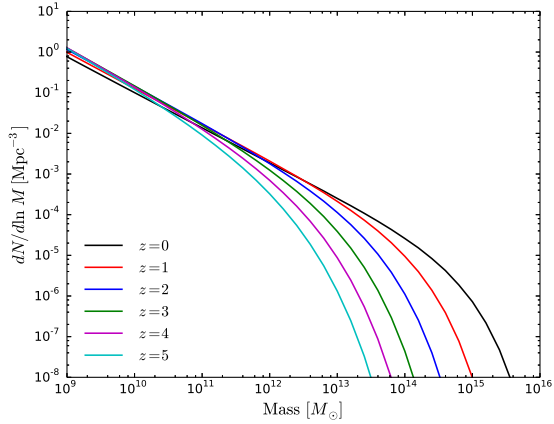}
}
\caption{Dark matter halo mass function $dN/d\ln M$ at $z=0,1,2,3,4,5$.
Here, we use the fitting formula from \citet{ref3}.}
\label{halomassfunction}
\end{figure}
%%%%%%%%%%%%%%%%%%%%%%%%%%%%%%%%%%%%%%%%%%%%%%%%%%%%%%%%%%%%%%%%%%%%%%5

{\ccminor Using the formalism of \cite{press1974formation}}, the halo mass function, the number of 
dark matter halos per unit comoving volume contained within the {\ccminor logarithmic mass interval} 
$d\ln M$, is given by
\begin{linenomath*}
\begin{equation}
	\frac{dN}{d{\rm ln}M}=\frac{\bar\rho}{M}
f(\nu)\frac{d{\rm ln}\sigma_M^{-1}}{d{\rm ln}M},
\label{eq:halodens}
\end{equation}
\end{linenomath*}
with the background matter density  {\ccmajor ${\bar\rho=\Omega_{m,0}\rho_{cr}}$ 
where $\rho_{cr}=(3H^2/8\pi G)$ is the critical density, and the variance $\sigma_M$ of the 
linear density contrast $\delta \equiv (\Delta \rho /{\bar \rho})$ is smoothed over the scale
$R=\left(\frac{3M}{4\pi\bar{\rho}}\right)^{1/3}$:}
\begin{linenomath*}
\begin{equation}
	\sigma_M^2=\int \frac{k^2dk}{2\pi^2} P(k)|\hat W(kR)|^2 \,.
\label{eq:massvar}
\end{equation}
\end{linenomath*}
Here, $P(k)$ is the linear matter power spectrum we calculate 
%by using {\rm CAMB} 
{ following} \citet{camb}, and $\hat W(kR) = 3j_1(kR)/kR$ for a top-hat filtering 
function. The significance $\nu=\delta_c/\sigma_M$ is related to the linear
critical density $\delta_c$ above which virialized halos can 
form\footnote{In some references, e.g. in \cite{ref3}, 
$\nu=(\delta_c/\sigma_M)^2$ is used instead of our definition here.}.
In the spherical collapse model, for example, a spherical region of radius $R$
collapses and virializes at redshift $z$ when the smoothed linear 
overdensity $\delta_R(\bm r, z)$ exceeds $\delta_{c,0}\approx 1.686$.
In the flat $\Lambda$CMD Universe, the linear growth factor, the time evolution of
the linear density contrast, is given by 
\begin{linenomath*}
\begin{equation}
\begin{split}
   {\mathcal D}(z)=\frac{\delta_c(z)}{\delta_{c,0}}&\propto\frac{5}{2}\Omega_{m,0}\sqrt{\Omega_{m,0}(1+z)^3+1-\Omega_{m,0}}\\
	&\times\int_{z}^{\infty}\frac{1+z'}{{[\Omega_{m,0}(1+z')^3+1-\Omega_{m,0}]}^{3/2}}dz'.
	\end{split}
\label{eq:Dz}
\end{equation}
\end{linenomath*}
We normalize ${\mathcal D}(z)$ to be unity at $z=0$.

For the multiplicity function $f(\nu)$, we adapt the Sheth-Tormen \citep{ref3} 
fitting formula, expressed in the form
\begin{linenomath*}
\begin{equation}
	f_{S-T}(\nu)=A\sqrt{\frac{2a}{\pi}}\left[1+(\nu^2a)^{-p}\right]\nu\exp{\left[-\frac{a\nu^2}{2}\right]},
\end{equation}
\end{linenomath*}
{\ccminor with parameters} $A=0.3222$, $a=0.707$, and $p=0.3$ that provide the 
the best fit {\ccminor{to numerical N-body simulations}} \citep{ref6,ref4,ref5}.
We show the resulting mass function $\frac{dN}{d{\rm ln}M}$ for
redshifts between $z=0$ and $z=5$  in figure \ref{halomassfunction}. 
Our mass function slightly underestimates that from the N-body simulations at high masses of $\gtrsim10^{15}~M\odot$, but 
it does not affect the main results of this paper. 
We shall use the mass functions in the following sections 
to estimate the redshift evolutions of galactic radius, gas density and shock velocity.

%%%%%%%%%%%%%%%%%%%%%%%%%%%%%%%%%%%%%%%%%%%%%%%%%%%%%%%%%%%%%%%%%%%%%%%%%%%%
\section{Merger rate and cosmic-ray luminosity density}
\label{sec:CR}
%%%%%%%%%%%%%%%%%%%%%%%%%%%%%%%%%%%%%%%%%%%%%%%%%%%%%%%%%%%%%%%%%%%%%%%%%%%%

{\ccmajor In this section, we calculate the CR input rate due to galaxy and halo mergers by using the halo 
mass function we have obtained in \refsec{halo}, and we estimate the energy converted into CRs from 
shocks in the gas component of the merging halos as follows}.

%The diffusive shock velocity of galaxy mergers is roughly in the range $v_s\approx3-9\times10^7cm/s$. 
%With this velocity, the CR acceleration efficiency is estimated to be $\eta\approx0.1$. 
%
There are three time scales characterizing the CR acceleration due to galactic halo mergers: 
the age of the Universe $t_{\rm age}$, 
the halo merger time $t_{\rm merger}$ {\ccminor which corresponds to the average time required to undergo one merger}, 
and the CR injection time (that is the shock-crossing time) $t_{\rm dyn}$,
which are, for a merger that happens at redshift $z$, given by
\begin{linenomath*}
\begin{equation}
	\begin{split}
		t_{\rm age}&=\int_z^\infty\left|\frac{dt}{dz'}\right|dz'\\
		t_{\rm merger}&=\left[\int d\zeta\, \frac{dN_m}{dzd\zeta}\left|{\frac{dz}{dt}}\right|\right]^{-1}\\
		t_{\rm dyn}&=\lambda\frac{R_{g}(z)}{v_s(z)}\,. \\
	\end{split}
\label{eq:times3}
\end{equation}
\end{linenomath*}
{Here, $|dt/dz| = 1/[(1+z)H(z)]$,
$dN_m/dzd\zeta$ \citep[given by][]{ref10,ref11} is the dimensionless 
merger rate per redshift interval $dz$ and per unit halo mass ratio $\zeta$, 
$R_g(z)$ is the mean galaxy radius, and $\lambda\sim 1$ parametrizes the 
orientation and geometrical uncertainty of the galaxy merger. 
%Since some halos may not merge during the age of the universe, 
With these time scales, the probability that a halo with mass $M$ experiences 
merger within the age of the universe is given by 
$P(M,z)=\exp(-t_{\rm merger}/t_{\rm age})$. 
Hence, assuming that the CRs are mainly protons, the {\it comoving} CR 
energy input rate per the logarithm of the CR energy $\varepsilon_p$ is
\begin{linenomath*}
\begin{equation}\begin{split}
	\varepsilon_pQ_{\varepsilon_p}(z)=\frac{E_{\rm merger}}{t_{\rm age}\mathcal C}=&
\epsilon_p \mathcal C^{-1}
\int_{M_{\rm min}}^{M_{\rm max}} dM\\
&\left[\frac{1}{2}\xi_{g}(M,z) Mv_s^2\right]\frac{dN}{dM}\frac{P(M,z)}{t_{\rm age}},
\end{split}
\label{eq:crinput}
\end{equation}
\end{linenomath*}
where $\xi_g(M_h,z)=M_{\rm gas}/M_h$ is the mass fraction in gas form, $\epsilon_p$ is the CR energy fraction (nominally taken as 0.1) 
and $\mathcal C=\ln(\varepsilon_p^{\rm max}/\varepsilon_p^{\rm min})$ is the normalization factor for a standard flat CR 
spectrum $N(\varepsilon_p) \propto \varepsilon_p^{-2}$. 
For $z\sim 1$, the typical maximum energy, $\varepsilon_p^{\rm max}$, is $\sim 10^{17}$ eV and ${\mathcal C} \simeq 18.4$ \citep{kashiyama2014galaxy}. 
However, $\varepsilon_p^{\rm max}$ varies with redshift, as we discuss in the next section.

%%%%%%%%%%%%%%%%%%%%%%%%%%%%%%%%%%%%%%%%%%%%%%%%%%%%%%%%%%%%%%%%%%%%%%%%%%%%
\subsection{Gas-mass fraction $\xi_g(M,z)$}
%%%%%%%%%%%%%%%%%%%%%%%%%%%%%%%%%%%%%%%%%%%%%%%%%%%%%%%%%%%%%%%%%%%%%%%%%%%%
The gas-mass fraction $\xi_g$ of dark matter halos depends on the star 
formation rate (SFR) and on the stellar mass $M_*=\chi_*(M_h,z)M_{h}$.
Here, we obtain $\chi_*=M_*/M_{h}$ from the $M_*(M_h)$ function inferred from
{\ccminor observations} by \cite{behroozi2013average}\footnote{In \cite{behroozi2013average}, the $M_*-M_h$ relation from $z=0-8$ is parameterized by equation (3). Here, we extend the domain of that function to $z=10$ {\ccminor considering that the uncertainty from high-redshift contributions is small.}}. 
We also {\ccmajor use the gas fraction} in normal galaxy $f_{g}=M_{\rm gas}/(M_{\rm gas}+M_*)$\footnote{%
In \cite{sargent2014regularity}, the gas fraction is written as 
$f_{\rm mol}$ instead.} 
measured in \cite{sargent2014regularity}. 
Combining the two observational results, we have constructed the redshift evolution of the 
gas-mass fraction in dark matter halos. That is, the gas-mass fraction $\xi_g$ is related to $f_{g}$ through 
$\xi_{g}^{\rm evo}=M_{\rm gas}/M_{h}=\frac{\chi_*f_{g}}{1-f_{g}}$, and using Eq. (26) in \cite{sargent2014regularity}, we find that 
%$\xi_{g}^{\rm evo}$ can be explicitly written as
%
\begin{linenomath*}
\begin{equation}
\xi_{g}^{\rm evo}=\chi_*\frac{f_{g}}{1-f_{g}}=\chi_* \frac{K}{M_*^{1-\beta'}}{\rm sSFR}^{\beta'}
\label{eq:xig}
\end{equation}
\end{linenomath*}
where $K=10^{\alpha_{\rm SFR}}$ is a constant and {the quantity ${\rm sSFR}$ (specific star formation rate) is the star formation rate per unit galaxy stellar mass.} For the gas fraction in normal 
galaxies we use the parameters $(\alpha_{\rm SFR},\beta')= (9.22\pm0.02,0.81\pm0.03)$, 
together with the expression for $\rm sSFR$ given in the appendix of \cite{sargent2014regularity}. 
In Fig.~\ref{fig:xigvs} (see red curves), we show the redshift evolution of the mean gas-mass fraction,
%A plot of the redshift dependence of the mean $\xi_{g}^{\rm evo}(z)$, which is defined through
\begin{linenomath*}
\begin{equation}
\langle\xi_g^{\rm evo}\rangle=\frac{\int\xi_g^{\rm evo}\frac{dN}{dM}dM}{\int\frac{dN}{dM}dM},\,
\end{equation}
\end{linenomath*}
as well as the constant gas fraction, $\xi_g=0.05$.

In our calculation, we take the lower and upper limit of the integration in Eq. (\ref{eq:crinput}) 
as $M_{\rm min}=10^{10}~\rm M_\odot$ and $M_{\rm max}=10^{15}~\rm M_\odot$, respectively. 
{\km There are two main reasons to choose the lower bound $10^{10}~\rm M_\odot$. 
First, considering the applicability of the $M_*(M_h)$ relation from \cite{behroozi2013average} and the gas fraction function (Eq.~\ref{eq:xig}), it is safe to truncate the halo mass at $M_h\sim10^{10}~\rm M_\odot$. Typically, dwarf galaxies reside in halos with mass less than $10^{10}~\rm M_\odot$ and we only have the constraints from observations at $z\simeq 0$. In our model, we consider the contribution from galaxy mergers up to the redshift $z=10$ where the  $M_*(M_h)$ function is not well tested for the lower halo masses. Also, the gas fraction function (Eq.~\ref{eq:xig}) is modeled from (normal) star-forming galaxies \citep{sargent2014regularity} and may not be valid for dwarf galaxies. 
Second, we estimate the low-mass halo contribution to CR luminosity density by extending the lower 
bound to $10^8~\rm M_\odot$ and found that the the contribution from $10^8-10^{10}~\rm M_\odot$ halos is 
$\lesssim 10\%$ of the total luminosity density in the low redshift ($z\lesssim3$), which implies that 
the conclusion of this paper does not depend sensitively on the mass range.}
\begin{figure}\centering
        \includegraphics[width=0.45\textwidth]{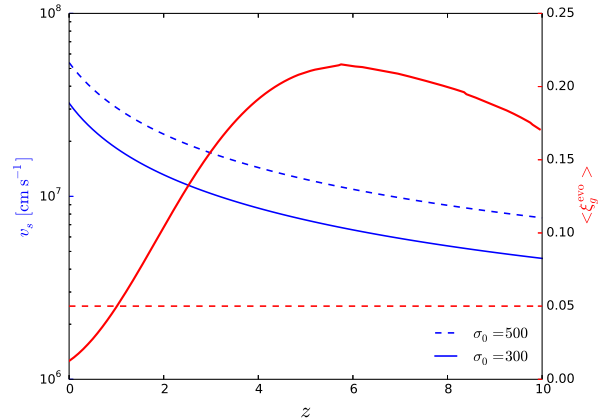}
        \caption{Redshift-dependent gas fraction $\xi_g(z)$ (red solid line) compared to
         a constant gas fraction $\xi_g=0.05$ (red dashed line) 
         for redshift-dependent shock velocity with $\sigma_0=300$ (blue solid line) and
         with $\sigma_0=500$ (blue dashed line), respectively.}
\label{fig:xigvs}
\end{figure}
%Later, we will discuss how $\xi_{g}=0.1$ and $\xi_{g}^{\rm evo}$ influence the results.

%%%%%%%%%%%%%%%%%%%%%%%%%%%%%%%%%%%%%%%%%%%%%%%%%%%%%%%%%%%%%%%%%%%%%%%%%%%%
\subsection{Shock velocity $v_s$}
%%%%%%%%%%%%%%%%%%%%%%%%%%%%%%%%%%%%%%%%%%%%%%%%%%%%%%%%%%%%%%%%%%%%%%%%%%%%
In the hierarchical clustering of large-scale structure scenario, the 
galactic-size {\ccminor halos are contained} inside larger cluster-size halos. 
The peculiar velocities of the galactic-size halos are, therefore, of order 
of the virial velocity of the cluster-sized halo. Here, we approximate the 
shock velocity of the galaxy merger from the pairwise velocity dispersion 
projected along the line of approach of two galaxies. For galaxies with a 
luminosity $L\approx L^*$ (where $L^*$ is the characteristic luminosity), 
\cite{davis1983survey} showed that the two-point correlation function at $r<20h^{-1}~\rm Mpc$ can be approximated by a power law
\begin{linenomath*}
\begin{equation}
\xi(r)=\left(\frac{r}{r_0}\right)^{\gamma},
\end{equation}
\end{linenomath*}
where $\gamma\approx1.7$ and $r_0\approx 5 h^{-1}~\rm Mpc$ is the correlation length, inside which 
galaxies are strongly correlated. Combining the hierarchical form of the three-point correlation 
function of galaxies \citep{groth1977statistical} and the cosmic virial theorem derived from 
the Layzer-Irvine equation, the collision (or shock) velocity can be written as
\begin{linenomath*}
\begin{eqnarray}  
v_s&=&\sqrt{\bar\sigma^2(r)}\nonumber\\
&\simeq&\sigma_0\left(\frac{r_0}{5h^{-1}~\rm Mpc}\right)^{\gamma/2}\left(\frac{r}{1h^{-1}~\rm Mpc}\right)^{-\gamma}\ {\rm km\ s^{-1}}.\,\,\,
\label{eq:vshock}
\end{eqnarray}
\end{linenomath*}
Given the average density of halos, we can estimate the average separation of galaxies through
\begin{linenomath*}
\begin{equation}
r=\left(\int \frac{dN}{dM}dM\right)^{-1/3}.
\label{eq:separation}
\end{equation}
\end{linenomath*}
%To find the redshift dependence of $r_0$, we assume that $r_0$ is proportional to the radii of 
%clusters and clusters at redshift $z$ will maintain clustered in the subsequent evolutions, in 
%other words, the mass remains the same. Therefore, 
%\begin{equation}
%r_0^{\rm same\ mass}\approx5h^{-1}{\rm Mpc} \left[\frac{\rho_{cl}(z=0)}{\rho_{cl}(z)}\right]^{1/3},
%\end{equation}
%and the density ratio will be given in the next section. However, considering the collision of 
%galactic clusters, this approximation may not be sufficiently accurate to describe the redshift 
%evolution of $r_0$. 
%
For our calculation it is necessary to consider the redshift dependance of the 
correlation function $\xi(r)$ in the nonlinear regime.
%in both the linear and nonlinear regime.
As a useful approximation, we adopt the stable clustering (SC) hypothesis
{\citep{peebles1974gravitational,davis1977integration}}, in which 
only the size (or separation between structures) of the clusters changes in 
time while the internal density structure of clusters stays intact.
This leads to $\xi(r,z)\propto(1+z)^{\gamma-3}$ and 
$r_0\propto(1+z)^{-(3-\gamma)/\gamma}$. 
Note that we only need the evolution of the nonlinear scale $r_0$, which 
is defined by $\xi(r_0,z)=1$. 
%This is why the redshift dependence of $\xi(r,z)$ in the linear regime is not required. 
A more accurate treatment %Hamilton et al. (1991; HKLM) 
\citep{hamilton1991reconstructing} describes the evolution of $\xi(r,z)$ from the linear to the 
nonlinear regime, and this treatment was generalized by 
\cite{peacock1996non} {\ccminor and by \cite{smith/etal:2003} %Peacock \& Dodds (1996) 
using another formula} for the nonlinear function. This generalized method gives $\xi(r,z)$ in 
the quasilinear regime and confirms that $\xi(r,z)\propto(1+z)^{\gamma-3}$ is valid in the nonlinear 
limit as well. 
Therefore, in this paper, we use $r_0^{\rm SC}\propto(1+z)^{-(3-\gamma)/\gamma}$ in 
Eq. (\ref{eq:vshock}) to find the shock velocities. 
We show the redshift dependence of the shock velocity $v_s(z)$ in Fig.~\ref{fig:xigvs} (see blue curves).

Note that, in our approximation of the shock velocity [Eq.~(\ref{eq:vshock})], 
the galaxy separation $r$ given by Eq. (\ref{eq:separation}) is overestimated, 
since this latter equation takes an average of the galaxies in a 
cosmological volume including clusters and voids. 
The mean separation of the galaxies in clusters, therefore, must be smaller than 
the value that we have adopted here, and this overestimate of $r$ would give a 
slight underestimate of $v_s$ inside clusters.
As a possible way to correct for this, we note that redshift surveys give $\sigma_0\sim500$ 
\citep{jing1998spatial,hawkins20032df,zehavi2002galaxy} as an average value for mergers in clusters.  
%(Jing et al, 1998; Hawkins et al, 2003; Zehavi et al, 2002)
%however, $\sigma_0$ should be larger if the separation given by equation 11 is used to determine 
%the mean separation of galaxies in a cluster. One method to lower the uncertainty is to scale 
%this effect into $\sigma_0$ using the local shock velocity $500km/s$ and local separation $r(z=0)$ 
%from equation 11. With these features, we obtain $\sigma_0\approx600$ 
A qualitatively appropriate correction for the cluster shock velocity may be obtained by scaling up $\sigma_0$ in
Eq. (\ref{eq:vshock}) from the local average value, concluding that a realistic value of
$\sigma_0$ for clusters is in the range $500\lesssim\sigma_0\lesssim600$.
{\km
In terms of rates, most galaxy mergers occur in the smaller mass halos containing fewer galaxies, as opposed to large clusters.
Considering the observational and theoretical uncertainties, we expect that the values of $\sigma_0$ lie in 
the range of $100\lesssim\sigma_0\lesssim1000$, and we take $300\lesssim\sigma_0\lesssim500$ as fiducial values.}

\section{Neutrino and $\gamma$-ray Production}
\label{sec:nugam}

Since in our model we need to consider the neutrino/$\gamma$-ray production rate up to redshift $z=10$, 
{\ccmajor we introduce here the redshift evolution function of the gas density, $g(z)=n(z)/n(z=0)$. 
To define this function we use the result that a sphere of gas will collapse and virialize once its 
density exceeds the value $1.686{\mathcal D}(z)^{-1}\rho_c(z)$ \citep{peebles1980large}. }
The mean density of the virialized gas is $\Delta_c\rho_c(z)$, where $\rho_c(z)=3H(z)^2/(8\pi G)$, and an
approximation of $\Delta_c(z)$ is $\Delta_c\approx178\Omega_m^{0.45}$ \citep{ref9}, where 
$\Omega_m=\Omega_{m,0}(1+z)^3/[\Omega_{m,0}(1+z)^3+1-\Omega_{m,0}]$. Since clusters are the largest 
virialized objects in the universe, we take $\rho_{\rm cl}(z)=g(z)n_{\rm cl,0}m_p\propto\Delta_c\rho_c(z)$, 
and we assume that galaxies, halos and clusters all share a universal $g(z)$,
\begin{linenomath*}
\begin{equation}
	g(z)=\frac{\Delta_c\rho_c(z)}{\Delta_{c,0}\rho_c(0)}=(1+z)^{1.35}[\Omega_{m,0}(1+z)^3+1-\Omega_{m,0}]^{0.55}.
\label{eq:fz}
\end{equation}
\end{linenomath*}
In the following sections, the relations $n_{\rm g}(z)=g(z)n_{\rm g,0},\ n_{\rm cl}(z)=g(z)n_{\rm cl,0}$ will 
be used {\ccmajor for the}  post-shock magnetic field and $pp$ optical depth.

\subsection{Galaxy mergers}
The maximum energy of CRs accelerated in the merger shocks will also evolve with $z$ due to the 
redshift dependance of the typical galactic radius and magnetic field characterizing the shocks. 
The magnetic field behind the shock in a galaxy merger is commonly parametrized as a fraction of the 
ram pressure \citep{kashiyama2014galaxy}, $B_s^2/8\pi=\frac{1}{2}\epsilon_Bn_{\rm g}m_pv_s^2\propto\rho v_s^2$.
%where $n_p\approx4n_g$ is the post-shock proton number density and $\xi_B\leq1$ is the ratio of the 
%post-shock magnetic to thermal energy.  
This implies a magnetic field
\begin{linenomath*}
\begin{eqnarray}
B_s&=&\sqrt{4\pi\epsilon_Bn_{\rm g,0}m_pg(z)v_s^2}\nonumber\\
&\simeq&14 \, \epsilon_{B,-2}^{1/2}n_{\rm g,0}^{1/2}g(z)^{1/2}\times\left(\frac{v_{s}}{300~{\rm km\ s^{-1}}}\right)\ {\rm \mu G}.
\label{eq:bshock}
\end{eqnarray}
\end{linenomath*}
The magnetic field in the disk region is expected to be higher than that in the halo region.  Although 
details depend on the geometry, for simplicity, we assume that a reasonably strong magnetic field is 
expected {\ccminor over scales between the galaxy radius $R_{\rm g}$ and a  gas scale height $h_{\rm g}$, 
which is taken as the characteristic scale height $h\sim{(3h_{\rm g}R_g^2/2)}^{1/3}$ in this work}. 
Then, the maximum CR energy is estimated to be \citep{drury1983introduction}%(Drury 1983)
\begin{linenomath*}
\begin{eqnarray}
\varepsilon_p^{\rm max}&\sim\frac{3}{20}eB_sh\frac{v_s}{c}\simeq1.3\times10^{16}~{\rm eV}\left(\frac{B_s}{30~\mu G}\right)\times\nonumber\\
&\left(\frac{h}{3~\rm kpc}\right)\left(\frac{v_s}{300~{\rm km\ s^{-1}}}\right).
\label{eq:epsilonmax}
\end{eqnarray}
\end{linenomath*}
The CRs are advected to the far downstream, and produce neutrinos and $\gamma$-rays during the advection. 
In reality, one needs to calculate neutrinos and $\gamma$-rays from the post-shock region especially when 
the $pp$ optical depth in the CR acceleration region is dominant. The emissions occur during 
$t_{\rm dyn}\sim h/v_s\simeq9.8~{\rm Myr}~(h/3~{\rm kpc})(300~{\rm km}~{\rm s}^{-1}/v_s)$. 
In this work, for simplicity, we take the CR reservoir limit, in which the CRs mostly escape into the ISM and the 
neutrino and $\gamma$-ray production mainly occurs in the ISM.  

After the CRs are accelerated in the shock, they will propagate in the host galaxy and cluster. In this 
process, neutrinos and $\gamma$-rays are generated from pions produced in inelastic $pp$ collisions. 
The meson production efficiency is $1-\exp(-f_{pp})$ where $f_{pp}=c\kappa_{pp}\sigma_{pp}g(z)\sum n_{i,0}t_i$ 
is the effective $pp$ optical depth. In this expression, $n_{i,0}$ is the local gas density of the medium, 
e.g. galaxies and clusters, $\sigma_{pp}=\sigma_{pp}(\varepsilon_p)$ is the $pp$ cross section given by 
\cite{kafexhiu2014parametrization}, $\kappa_{pp}=0.5$ is the inelasticity coefficient and $g(z)$ (see Eq. (\ref{eq:fz}))
represents the redshift evolution of the gas density.

Let us consider galaxies that are merging at $z$. Inside the merged galaxy, $f_{pp}^{\rm g}$ is determined by the 
time spent by the CRs undergoing $pp$ collisions, which depends on the CR injection time and and the diffusion 
time in the medium.  The dynamical time is given by the third of Eq. (\ref{eq:times3}), while the diffusion time is 
$t_{\rm diff}=h(z)^2/(6D_{\rm g})$, where $h(z)$ is the effective gas size at $z$ and $D_g$ is the diffusion 
coefficient {\km in the galactic ISM gas}.  Here, we use a combined large and small angle diffusion expression 
as in \cite{senno2015extragalactic}, $D=D_c[(\varepsilon/\varepsilon_{c,g})^{1/2}+(\varepsilon/\varepsilon_{c,g})^2]$, 
where $D_c=cr_L(\varepsilon_{c,g})/4$ and $\varepsilon_{c,g}$ is determined from $r_L(\varepsilon_{c,g})=l_c/5$. 
Here, $r_L$ and $l_c$ are the Larmor radius and coherence length {in the galaxy environment} respectively. 
For local normal galaxies, the gas density in the disk is $n_{\rm g,0}\sim 1~{\rm cm}^{-3}$, whereas 
{\ccmajor the average density in the {galactic} halo is smaller}, $n_{\rm g,0}\sim0.1~{\rm cm}^{-3}$. 
The magnetic field of local normal galaxies is $\sim4~\mu G$ and that of  star-forming galaxies is $\sim6~\mu G$, 
respectively \citep{ref7,ref8}.  For the density and magnetic field of merging galaxies, we take values higher 
than those of normal galaxies, since the galaxies may enter the starburst phase during the merger. 
Specifically we adopt {a mean value} $n_{\rm g,0}=1~{\rm cm}^{-3}$. Thus, we have
\begin{linenomath*}
\begin{equation}
	t_{\rm diff} \simeq 3.2\times 10^5~{\rm yr}\ \left(\frac{h(z)}{3~{\rm kpc}}\right)\left[(\varepsilon/\varepsilon_{c,g})^{1/2}+(\varepsilon/\varepsilon_{c,g})^2\right]^{-1}
\label{eq:tdiff}
\end{equation}
\end{linenomath*}
where
\begin{linenomath*}
\begin{equation}
\varepsilon_{c,g}\simeq 1.7\times10^9~{\rm GeV}\ \left(\frac{h(z)}{3~{\rm kpc}}\right)\left(\frac{B_g}{30~\mu G}\right).
\label{eq:epsilonc}
\end{equation}
\end{linenomath*}

Calculations of the neutrino and $\gamma$-ray emission depend on details of the spatial extension and time evolution of the shock region and its surrounding environment. The latter is also modified by the shock, star-formation, and outflow. 
For simplicity, we treat a double-galaxy merger system as one CR reservoir for the injection 
by the merger shock, which {\ccminor is conservative} since there should also be the emissions from the accelerator. {\ccminor  A similar treatment} for neutrino sources with active accelerators is used in the galaxy cluster model \citep{murase2008cosmic,fang2018linking}. 
Then, the effective $pp$ optical depth is estimated to be $f_{pp}^{\rm g}=\kappa_{pp} cg(z)n_{\rm g,0}\sigma_{pp}{\rm min}[{t_{\rm dyn},t_{\rm diff}}] \simeq 0.24~g(z)
\left(\frac{n_{\rm g,0}}{1~\rm cm^{-3}}\right) \left(\frac{\sigma_{pp}}{50~\rm mb}\right)
\left(\frac{{\rm min}[t_{\rm dyn},t_{\rm diff}]}{10~\rm Myr}\right)$ in the merging galaxy system. 
The ambient magnetic field energy may be taken to be a fraction of the merging galaxy system's virial energy, as 
$B_{\rm g}^2 R_{\rm g}^3 \propto GM_{\rm g}^2/R_{\rm g}$, i.e. $B_{\rm g} \propto \rho_{\rm g} R_{\rm g} \propto g(z) R_{\rm g}(z)$. 
 
The typical galactic radii evolve with redshift $z$, and considering the merger history of galaxies, 
it is apparent that the mean radii of galaxies at $z$ should be smaller than $R_{\rm g,0}/(1+z)$, where 
$R_{\rm g,0}\approx10~{\rm kpc}$ is the radius of local Milky Way-like galaxies. \cite{shibuya2015morphologies} %T.Shibuya et al.
studied the redshift evolution of the galaxy effective radius $r_e$ using $Hubble\ Space\ Telescope$ (HST) 
samples of galaxies at $z=0-10$, finding $r_e\propto(1+z)^{-1.0}-{(1+z)}^{-1.3}$ with $r_e\propto (1+z)^{-1.10\pm0.06}$ 
as a median. Hence, in this paper, we assume that the average galaxy radius evolves with respect to $z$ 
as $R_{\rm g}=R_{\rm g,0}(1+z)^{-1.10}$.

As for the scale height $h_g(z)$, based on the surface photometry analysis of edge-on spiral galaxies,
e.g. NGC 4565, 
it has been shown that the {\ccminor scale height} of gas in local disk galaxies is approximately 
$h_{g,0}\approx300-400$ pc \citep{bahcall1980universe,van1981surface}. Later studies of NGC 891 
\citep{kylafis1987dust}, NGC 5097\citep{barnaby1992distribution} etc. also agree with this estimate. 
Considering that a merger can lead to entering a star-forming phase, we assume $h_{g,0}=500pc$ and 
we assume the same {\ccminor redshift dependence} as for $R_{\rm g}$, e.g. $h_{\rm g}(z)=(1+z)^{-1.10}h_{\rm g,0}$. 
Then we take $h={(3h_{\rm g}R_{\rm g}^2/2)}^{1/3}$.%, and $n_g$ is determined by $M_{\rm gas}=(4\pi/3)m_pn_g h^3$.} 
 
\subsection{Interactions in the host cluster and cluster mergers}
After escaping the galaxy, the CRs may continue to collide with the gas of the host cluster, where
$t_{\rm diff}=R_{\rm cl}(z)^2/(6D_{\rm cl})$. 
Here, we assume a magnetic field $B_{\rm cl,0}\approx1~\mu G$ with a coherence length $l_{c,\rm cl}\approx 30~\rm kpc$. 
This implies $\varepsilon_{c,\rm cl}\approx5.6\times10^9~\rm GeV$. 
For a cluster of mass $10^{15}~\rm M_\odot$, the virial radius is 
$R_{\rm cl,0}=(3M/(4\pi\rho_{\rm cl,0}))^{1/3}\approx2.1~\rm Mpc$. 
Since $R_{\rm cl}$ is the approximate boundary of clustered/correlated galaxies, it should have 
the same {{\ccminor redshift dependence}} as $r_{0}^{\rm SC}$. 
Using the stable clustering approximation, we obtain $R_{\rm cl}\propto (1+z)^{-(3-\gamma)/\gamma}$. 
Similarly, we can calculate the diffusion time in clusters as
$t_{\rm diff,cl}=1.2[(\varepsilon/\varepsilon_{c,\rm cl})^{1/2}+(\varepsilon/\varepsilon_{c,\rm cl})^2]^{-1}\ \rm Gyr$.
Assuming that the injection time of CRs ($t_{\rm inj}$) at redshift $z$ is the cluster age (of order the Hubble time)
$t_{\rm age}(z)$, likewise we obtain the optical depth $f_{pp}^{\rm cl}=\kappa_{pp} cg(z)n_{\rm cl,0}\sigma_{pp} 
{\rm min}[{t_{\rm inj},t_{\rm diff,cl}}]\simeq0.24~g(z)
\left(\frac{n_{\rm cl,0}}{{10}^{-3}~\rm cm^{-3}}\right) \left(\frac{\sigma_{pp}}{50~\rm mb}\right)
\left(\frac{{\rm min}[t_{\rm age},t_{\rm diff}]}{10~\rm Gyr}\right)$, where $n_{\rm cl,0}$, the intercluster gas 
density, is assumed to have the {\ccmajor typical value $ n_{\rm cl,0} \sim 10^{-4}-{10}^{-2}~\rm cm^{-3}$ 
\citep{croston2008galaxy}}, which can be higher in cooling core clusters.  
The magnetic field may also depend on $z$ as $B_{\rm cl}\propto \rho_{\rm cl} R_{\rm cl} \propto g(z) R_{\rm cl}(z)$.

{\km Halo mergers will also lead to galaxy group and galaxy cluster mergers, after some halos have
grown above a certain size which may be taken to be roughly of order $M_h \sim 10^{13}~\rm M_\odot$.
We simplify the calculations as follows. For low-mass mergers, we expect that the $pp$ interactions occur 
mainly in gas with an ISM density characteristic of galaxies, while for high-mass mergers the $pp$ 
interactions occur mainly in gas with an IGM density characterizing the intra-cluster gas. 
In addition, there will be a component of $pp$ interactions due to low-mass merger CRs which
escape from the colliding galaxy system into the IGM. 
Thus, we expect that the all-flavor neutrino production rate consists of a 
galaxy part $\varepsilon_\nu Q_{\varepsilon_\nu}^{(\rm g)}$ and a 
cluster/group part $\varepsilon_\nu Q_{\varepsilon_\nu}^{(\rm cl)}$} plus a
weaker galaxy-cluster term,
\begin{linenomath*}
\begin{equation}\begin{split}
\varepsilon_\nu Q_{\varepsilon_\nu}^{(\rm g)}= &\frac{1}{2}(1-e^{-f_{pp}^{\rm g}})\varepsilon_pQ_{\varepsilon_p}^{\rm (LM)}\\
 \varepsilon_\nu Q_{\varepsilon_\nu}^{(\rm cl)}=& \frac{1}{2}[ (1-e^{-f_{pp}^{\rm cl}}) \varepsilon_pQ_{\varepsilon_p}^{\rm (HM)}\\
& +\eta(1-e^{-f_{pp}^{\rm cl}}) e^{-f_{pp}^{\rm g}} \varepsilon_pQ_{\varepsilon_p}^{\rm (LM)}],
\end{split}
\label{eq:neu_input}
\end{equation}
\end{linenomath*}
where the energies of the neutrinos and CR protons are related by $\varepsilon_\nu\approx0.05\varepsilon_p$. Note that the luminosity density evolution of neutrinos and $\gamma$-rays is different from that of CRs in general. 
{\km 
In the first line of Eq. \ref{eq:neu_input} $\varepsilon_pQ_{\varepsilon_p}^{(\rm LM)}$ is the CR input rate
(see Eq.~\ref{eq:crinput}) from galaxy mergers in low-mass (LM) halos, e.g. 
$[10^{10}~{\rm M_\odot},10^{13}~\rm M_{\odot}]$. 
In the second line $\varepsilon_pQ_{\varepsilon_p}^{(\rm HM)}$ is the CR input rate of the high-mass (HM) halo 
mergers, in the interval $[10^{13}~{\rm M_\odot},10^{15}~\rm M_{\odot}]$.
The factors $\frac{1}{2}(1-e^{-f_{pp}^{\rm i}}) \varepsilon_pQ_{\varepsilon_p}^{\rm (j)}$ are the neutrino luminosity density 
from CRs originating from mergers of mass $(j)$ in gas of density $i$.
For our fiducial parameters, these two components constitute the largest fraction of the neutrino budget. 
Nevertheless, for completeness, we have included in the third line of Eq. \ref{eq:neu_input} 
the sub-dominant effect due the CRs produced in 
galaxy mergers which may escape the host galaxies and collide with intra-cluster gas to produce neutrinos. 
(This can be important only if the $pp$ interactions in galaxy mergers are inefficient.)
We introduce a parameter $\eta$ to represent the fraction of galaxy mergers that occur inside clusters which
lead to some CRs escaping into the gas halo. This can occur preferentially at lower redshifts. Since the boundary 
between LM and HM is ambiguous and the fraction $\eta$ can change with redshift, this parameter is very 
uncertain, and may conservatively be estimated as between 0.1 and at most 0.5. Fortunately, the contribution of 
this higher-order third component depending on $\eta$ is small compared to the first two components in Eq. 
\ref{eq:neu_input}, due to the factor $e^{-f_{pp}^{\rm g}}$. At $z=1$, the ratio between the third line and the 
first line is $\leq10\%$ even if $\eta$ is assumed to be unity, and it is increasingly negligible at higher 
redshift since $f_{pp}^{\rm g}$ increases as the gas density increases. Therefore, the exact value of $\eta$ 
does not significantly influence our final results.}

\section{Diffuse neutrino and $\gamma$-ray spectra}
With the above, we are able to determine the CR energy input rate, $\varepsilon_\nu Q_{\varepsilon_\nu}^{(\rm g)}$ and $\varepsilon_\nu Q_{\varepsilon_\nu}^{(\rm cl)}$. Figure \ref{fig:CRinput} shows the CR input power over the whole mass interval $10^{10}\rm M_{\odot}-10^{15}\rm M_{\odot}$ as a function of $z$ as well as the LM and HM components of $(\sigma_0=300,\ \xi_g^{\rm evo})$ scenario. 
As can be seen from the redshift distribution of the CR energy input, using a redshift evolving gas fraction $\xi_g^{\rm evo}$, a significant fraction of
this occurs at redshifts $z\gtrsim 3$, above which a significant $\gamma\gamma$ attenuation
of the accompanying high-energy $\gamma$ rays at $\gtrsim20-30$ GeV energies can be expected \citep[e.g.,][]{Chang+16grbnugam,Xiao+16nuhn}. 
In addition, from Figure \ref{fig:CRinput}, we find that the high-mass and low-mass components are comparable in local mergers, implying that the cluster/group merger contribution is also important. Also, the galaxy merger contribution to the CR luminosity density is more important at $z\gtrsim2$. 

\begin{figure}\centering
	\includegraphics[width=0.45\textwidth]{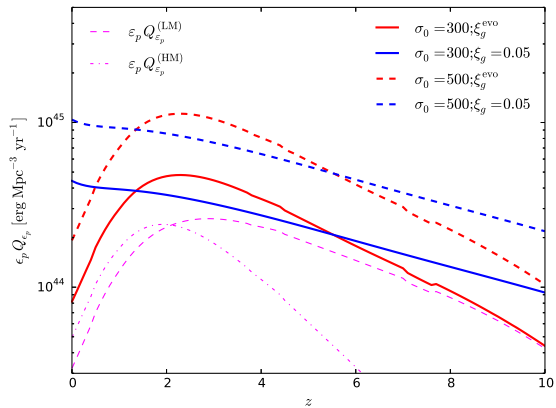}
	\caption{CR energy input rate versus redshift. 
        The red lines correspond to a redshift-dependent gas fraction $\xi_{g}^{\rm evo}$ 
        and the blue lines are for a redshift-independent gas fraction $\xi_g=0.05$, 
        while the solid lines are for $\sigma_0=300$ and the dashed are for  $\sigma_0=500$, 
        repectively. The dashed and dash-dotted magenta lines are LM and HM components of $(\sigma_0=300,\ \xi_g^{\rm evo})$ scenario. {\ccminor Here LM and HM denote the low-mass ($10^{10}~\rm M_\odot-10^{13}M_\odot$) and high-mass ($10^{13}~\rm M_\odot-10^{15}M_\odot$) intervals, respectively.}}
\label{fig:CRinput}
\end{figure}

Given the neutrino input rate, the all-flavor neutrino flux can be expressed as \citep{Murase:2015xka}
\begin{linenomath*}
\begin{equation}
	\varepsilon_\nu^2\Phi_{\varepsilon_\nu}=\frac{c}{4\pi}\int\frac{\varepsilon_\nu Q_{\varepsilon_\nu}^{(\rm g)}+\varepsilon_\nu Q_{\varepsilon_\nu}^{(\rm cl)}}{(1+z)}\left|\frac{dt}{dz}\right|dz ~,
\label{eq:diffnu}
\end{equation}
\end{linenomath*}
 Based on the branching ratio between charged and neutral pions, the {\it initial} diffuse $\gamma$-ray
energy spectrum is expected to be given by $\varepsilon_\gamma^2\Phi_{\varepsilon_\gamma}=
\frac{2}{3}\varepsilon_\nu^2\Phi_{\varepsilon_{\nu}}|_{\varepsilon_\nu=0.5\varepsilon_\gamma}$. Since however the high-energy 
$\gamma$ rays can annihilate with lower energy photons, such as those from the extragalactic background 
light (EBL) and the cosmic microwave background (CMB),  we introduce an attenuation factor 
$\exp[-\tau_{\gamma\gamma}(\varepsilon_\gamma,z)]$ to the integration, where $\tau_{\gamma\gamma}$ is 
the $\gamma\gamma$ optical depth at redshift $z$. In this paper, we use the optical depth provided by 
%Finke et al. (2010) and %Y. Inoue et al (2013)
\citep{finke2010modeling,inoue2013extragalactic} for low-redshift ($z\leq5$) and 
high-redshift ($z>5$) inputs, respectively. The attenuated $\gamma$-ray flux is then
\begin{linenomath*}
\begin{equation}\begin{split}
	\varepsilon_\gamma^2\Phi_{\varepsilon_\gamma}=\frac{c}{4\pi}\int\frac{2}{3}&\left[\frac{\varepsilon_\nu Q_{\varepsilon_\nu}^{(\rm g)}+\varepsilon_\nu Q_{\varepsilon_\nu}^{(\rm cl)}}{(1+z)}\left|\frac{dt}{dz}\right|\right]\\
	&\times\exp[-\tau_{\gamma\gamma}(\varepsilon_\gamma,z)]dz
\end{split}
\label{eq:diffgamaten}
\end{equation}
\end{linenomath*}
with $\varepsilon_p=10\varepsilon_\gamma(1+z)$. In addition, the electron-positron pairs produced in the
$\gamma\gamma$ annihilations will subsequently scatter off the ambient diffuse photon backgrounds,
leading to an electromagnetic cascade which in part compensates for the attenuation, while
reprocessing the photon energy towards lower energies, which can be detected by, e.g. the {\it{Fermi}}-LAT
instrument. In this paper, for simplicity, we use the universal form for the resulting cascaded $\gamma$-ray spectrum 
given by \cite{berezinsky1975cosmic} \citep[see also, e.g.,][]{murase2012constraining,senno2015extragalactic},
\begin{linenomath*}
\begin{equation}
\varepsilon_\gamma\frac{dN}{d\varepsilon_\gamma}\propto G(\varepsilon_\gamma)=
\begin{cases}
\left(\frac{\varepsilon_\gamma}{\varepsilon_\gamma^{\rm br}}\right)^{-1/2} & \varepsilon_\gamma\leq\varepsilon_\gamma^{\rm br}\\
\left(\frac{\varepsilon_\gamma}{\varepsilon_\gamma^{\rm cut}}\right)^{-1} &   \varepsilon_\gamma^{\rm br}<\varepsilon_\gamma<\varepsilon_\gamma^{\rm cut}\\
\end{cases}
\label{eq:gamcasc}
\end{equation}
\end{linenomath*}
where $\varepsilon_\gamma^{\rm cut}$ is defined by $\tau_\gamma(\varepsilon_{\gamma}^{\rm cut},z)=1$ 
and $\varepsilon_\gamma^{\rm br}=0.0085~{\rm GeV}(1+z)^2\left(\frac{\varepsilon^{\rm cut}_\gamma}{100\rm GeV}\right)^2$.

\begin{figure*}\centering
{\includegraphics[width=0.45\textwidth]{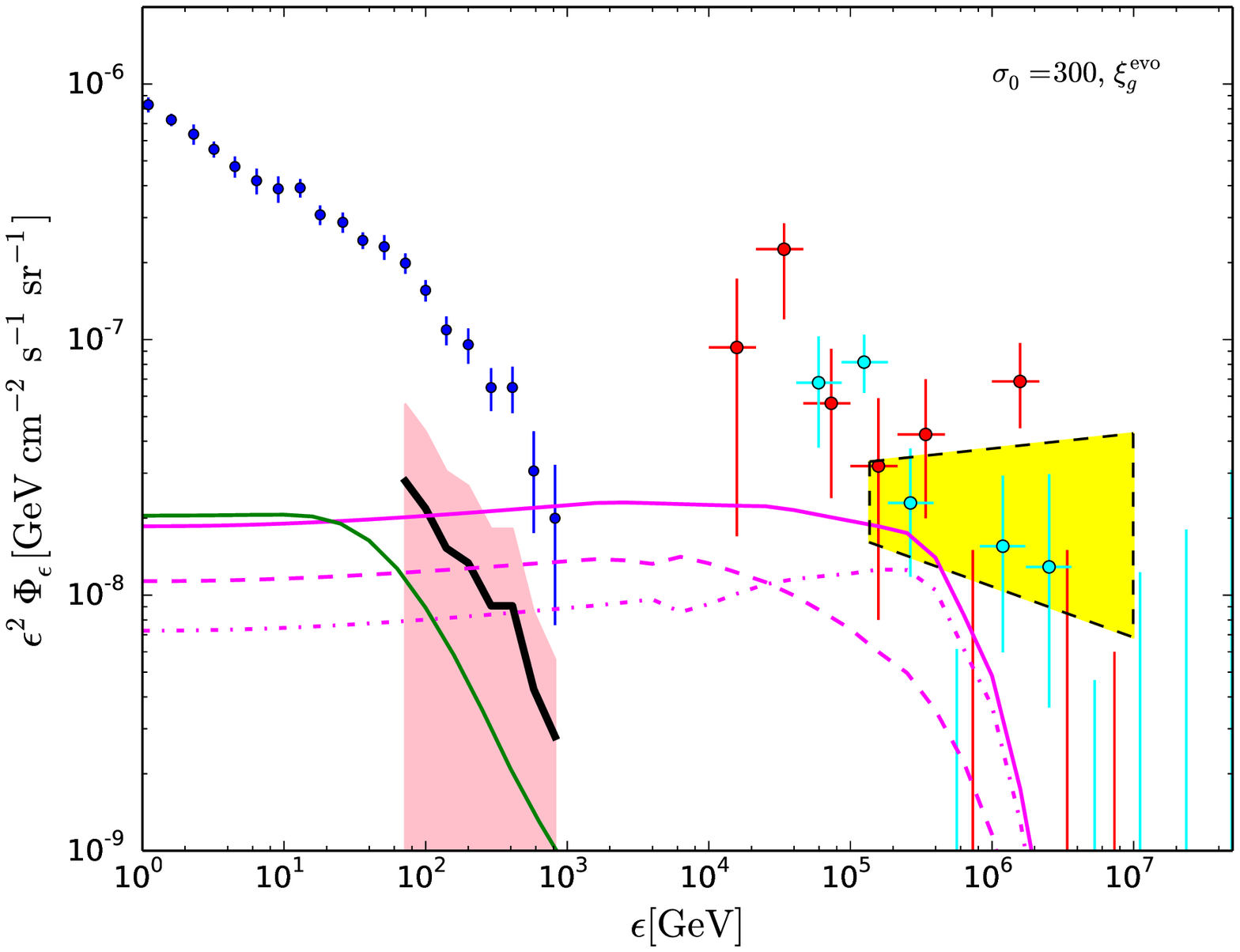}}
	{\includegraphics[width=0.45\textwidth]{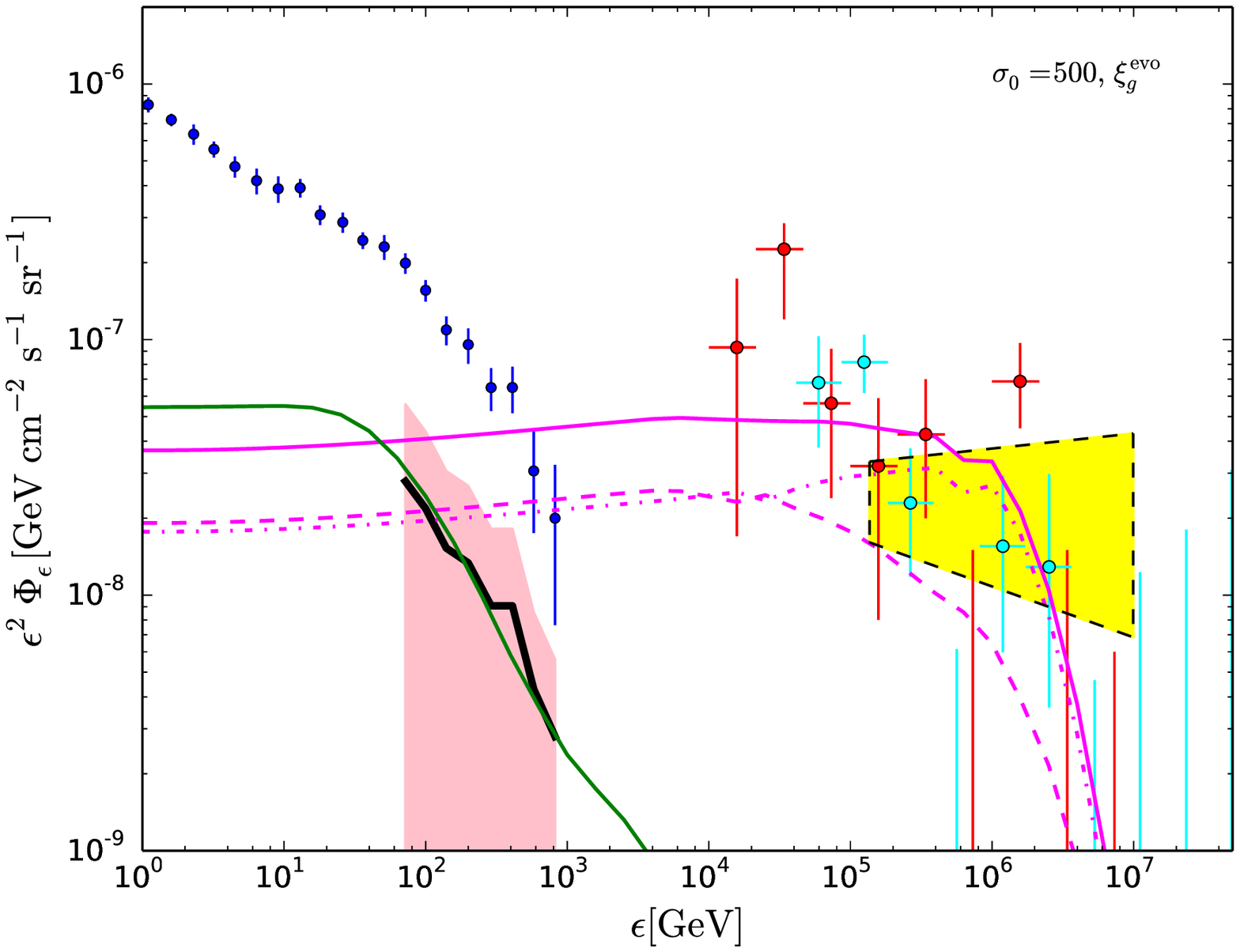}}	\caption{Left panel: Neutrino (all flavor) and $\gamma$-ray fluxes from halo mergers with redshift-evolving gas fraction $\xi_g^{\rm evo}$, $R_{g,0}=10~{\rm kpc},\ H_{g,0}=500$~pc. 
The shock velocity is obtained using $r_0^{\rm sc}(z)$ and $\sigma_0=300$. The magenta line is 
the neutrino spectrum while the green line is the corresponding $\gamma$-ray spectrum. Galaxy and cluster 
contributions to the neutrino flux are illustrated as the dashed and dash-dotted lines, respectively. 
Right panel: same as left panel except $\sigma_0=500$ is utilized for $v_s.$}
	\label{fig:const_gas_fraction}
\end{figure*}

\begin{figure*}\centering
	{\includegraphics[width=0.45\textwidth]{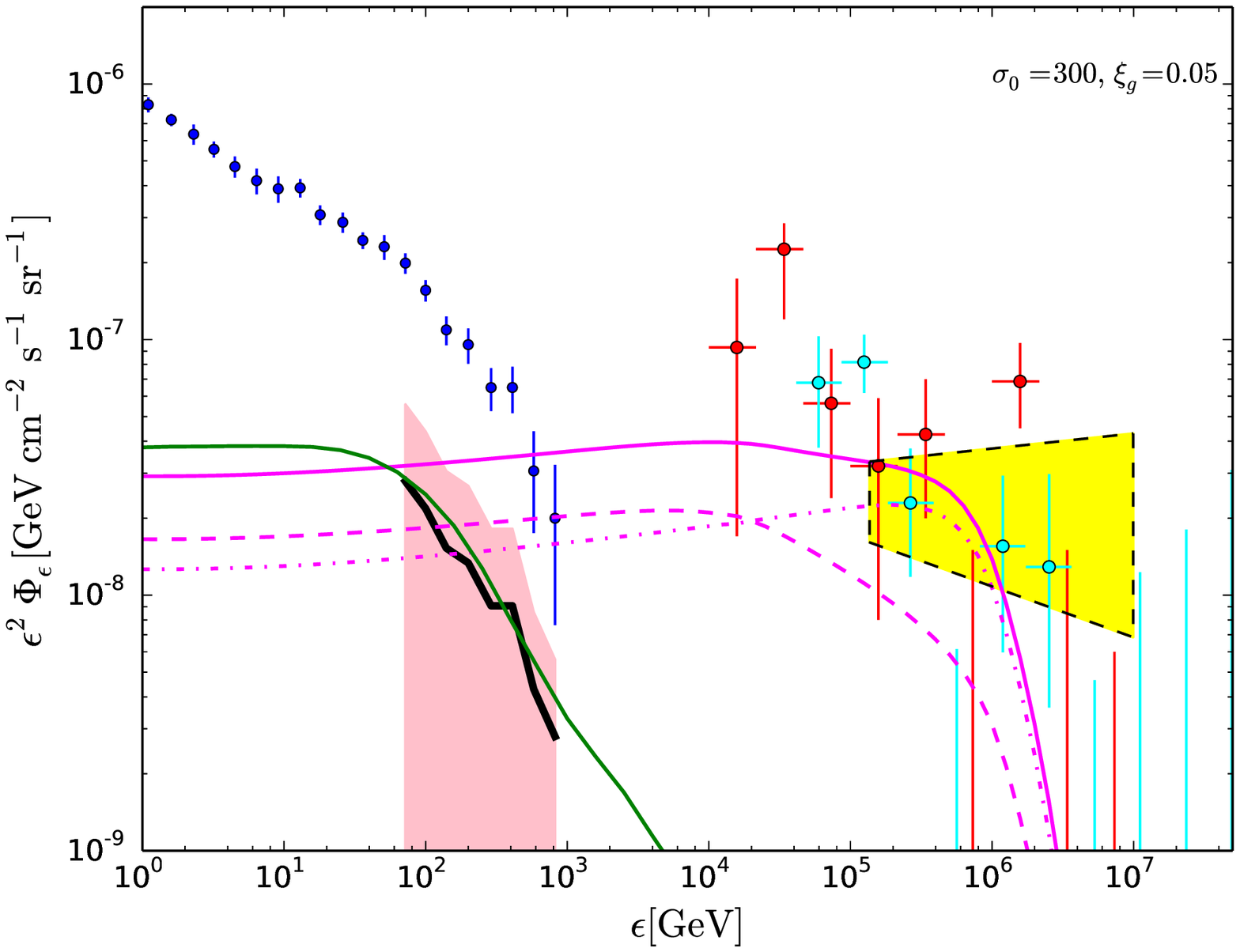}}
	{\includegraphics[width=0.45\textwidth]{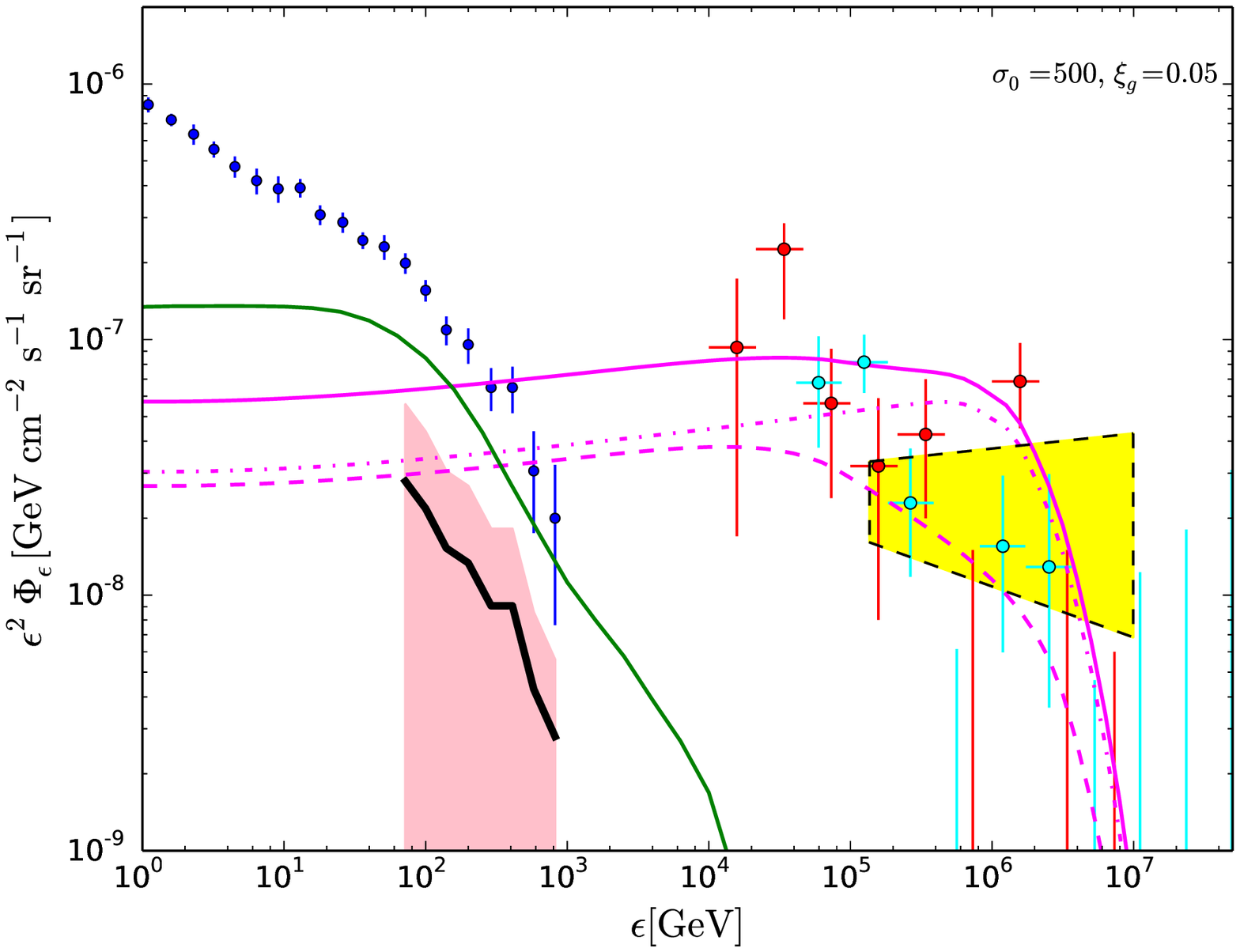}}
	\caption{Left panel: same as Fig.~\ref{fig:const_gas_fraction} (a), $\sigma_0=300$, except that $\xi_g=0.05$ is used 
to estimate the redshift evolution {\ccminor of the halo gas fraction}. Right panel: same as left figure except with $\sigma_0=500$.}
	\label{fig:evo_gas_fraction}
\end{figure*}

The all-flavor diffuse neutrino and $\gamma$-ray fluxes are plotted in Fig. \ref{fig:const_gas_fraction}, 
together with the IceCube observed astrophysical neutrinos.
{\ccminor The red points and cyan points correspond to the all-flavor averaged neutrino flux \citep{aartsen2015combined,aartsen2016observation} and the 6-year high energy starting-events (HESE)}~\citep{Aartsen:2017mau}, respectively. The {{\it Fermi}}-LAT observed {\ccminor total extragalactic $\gamma$-ray background (EGB)} \citep{ackermann2015spectrum} is shown by the blue points. The yellow area is the best-fit to the 
up-coming muon neutrinos {\ccminor scaled to three-flavor}. {\ccminor Figure \ref{fig:const_gas_fraction} shows} the results for an assumed redshift-dependent  gas fraction $\xi_g^{\rm evo}$, as illustrated in Fig. 
\ref{fig:const_gas_fraction}(a) for $\sigma_0=300$ and in Fig. \ref{fig:const_gas_fraction}(b) for 
$\sigma_0=500$, showing the effect of the corresponding different shock velocities $v_s$. 
In each figure, the magenta line represents the neutrino flux while the green line illustrates the 
corresponding $\gamma$-ray flux after cascading down. The galaxy and cluster contributions to the neutrino 
flux are plotted in dashed lines and dash-dotted lines.  The non-blazar \citep{Fermi16} component 
of the {\ccminor unresolved extragalactic gamma-ray background} is shown as the pink area. 

For illustration purposes, we consider next the corresponding results using the redshift-independent gas fraction $\xi_g=0.05$, 
which are shown in figures \ref{fig:evo_gas_fraction} (a) and \ref{fig:evo_gas_fraction} (b). The comparison 
between the galaxy and cluster components indicates that the high-energy neutrinos are dominantly 
produced by the propagation of CRs in the clusters. This is a consequence of the rapid redshift 
evolution of the galaxy radius, since the size of the host galaxy limits the maximum CR energy as well as the neutrino production efficiency by restricting the diffusion time. 
In addition, a redshift-dependent $\xi_g^{\rm evo}$  boosts the CR budget to a relatively higher redshift 
($z\approx3$), as can be seen from the red line in Fig.\ref{fig:CRinput}, which as was expected leads 
to a reduction in the $\gamma$-ray flux. From these figures, we can also see that even with the moderate 
sensitivity of the results to the parameters $r_0$ and $\sigma_0$, the results can broadly fit a significant 
fraction of the IceCube data without violating the non-blazar EGB. Conversely, the $\gamma$-ray and neutrino 
fluxes are significantly constrained in this scenario, indicating that the halo and galaxy mergers 
can be regarded as {\ccminor promising sources of neutrinos} in the context of multi-messenger studies. 

\section{Discussion}
\label{sec:dis}

In this work, we investigated the contribution of halo mergers to {\ccminor the diffuse 
neutrino} and $\gamma$-ray backgrounds, and tested whether the non-blazar diffuse $\gamma$-ray background 
{\it{Fermi}} {\ccminor constraint is violated}. 
Our results differ from previous work by \cite{kashiyama2014galaxy} in that we studied both galaxy and cluster/group
mergers out to higher redshifts, up to $z\approx10$, by considering the redshift evolution of the
average galactic radius, the shock velocity and the gas content inside the halos, as well as the 
galactic/intergalactic magnetic fields. 

The redshift evolution of galaxy radius implies that there exist more protogalaxies, or equivalently more 
mergers at higher redshift. In fact, the merger rate calculated using our approximate approach {\ccminor verifies 
this conjecture}, as well as being consistent with the Illustris simulations \citep{rodriguez2015merger}.
Also, our estimates of the gas fraction $\xi_{g}^{\rm evo}$ based on the correlation between the galactic 
gas content and the star formation rate shows that the gas in high-redshift halos is relatively denser 
than in the current epoch halos. The net effect is that high-redshift halo mergers can contribute 
{\ccminor a significant fraction of} the cosmic rays that are capable of producing high-energy neutrinos, as 
shown in Fig. \ref{fig:CRinput}. This is crucial since the accompanying $\gamma$-ray photons in the 
ensuing $pp$ collisions at high redshifts can be sufficiently absorbed via $\gamma\gamma$ annihilations 
against CMB and EBL photons. In both cases with $\xi_{g}^{\rm evo}$, our results indicate that high-redshift 
galaxy/halo mergers can explain a large fraction of the IceCube observed diffuse neutrinos up to 
{\ccmajor $\sim $PeV}, with an accompanying $\gamma$-ray diffuse observed flux which is below the 
non-blazar $Fermi$ constraints. 

{\km We note that according to our calculation, the diffuse flux of CRs that survive from energy losses via $pp$ collisions is less 
than $10^{-8} \ {\rm GeV\ cm^{-2}\ sr^{-1}\ s^{-1}}$, which is lower than the observed CR flux around the knee or sub-ankle energy.} 
%($\sim10^{-5}-10^{-4} \ {\rm GeV\ cm^{-2}\ sr^{-1}\ s^{-1}}$).} 

The CR acceleration efficiency $\epsilon_p$ is expected to be $\sim0.1$ based on the {\ccmajor diffusive shock 
acceleration theory}.  The redshift dependences of gas fraction $\xi_g^{\rm evo}$ and galaxy radius are relatively 
well-modeled from current theories and observations, so our scenario can put a tighter constraint on the shock 
velocity of galaxy mergers.  However, there are large uncertainties in the model. For example, the maximum energy 
depends on the magnetic field strength that is highly uncertain. On the other hand, the fiducial value of 
$\sim10$~PeV is not far from the knee energy at $\sim3$~PeV, so our assumption is reasonable. 
One of the most important uncertainties is caused by the shock velocity. Our fiducial parameters ($\sigma_0=300$ 
with $\xi_{g}^{\rm evo}$) imply a lower neutrino flux compared to the observations. This could be overcome 
by assuming a higher velocity with a stronger magnetic field. Or it may be possible to achieve the IceCube flux 
at $\gtrsim0.1$~PeV without exceeding the {\it{Fermi}} constraint by increasing the cluster contribution. 
However, the cluster contribution is more uncertain. Non-thermal emissions from merging/accreting clusters 
have been studied by various authors \citep[e.g.,][]{Berrington:2002bh,fujita2001nonthermal}. The Mach 
number of shocks {on the} cluster scales is so low due to the high temperature of the intra-cluster 
medium that the shock may not be strong enough to have a hard spectrum of $s\sim2$. 

{\km We note that, in addition to mergers, also cluster accretion shocks and powerful jets
from radio-loud AGNs can contribute to CR acceleration inside the clusters/groups, as considered in the 
{\ccminor previous literature} \citep[e.g.,][and references therein]{murase2013testing,fang2018linking}.  
One of the generic features of the CR reservoir scenario is that different possibilities for CR 
acceleration are not mutually exclusive, and additional contributions from various CR accelerators 
may enhance the neutrino flux. {\ccmajor Another CR source that can be associated with galaxy mergers 
is that the compression of the ISM gas can trigger an intense starburst. As discussed by
\citet{charbonnel2011star}, two processes in colliding galaxies could induce starburst: radial gas 
inflows can fuel a nuclear starburst, while gas turbulence and fragmentation can drive an extended starburst 
in clusters.} Such intense star-formation can naturally lead to the injection of CRs from the ensuing 
massive stellar deaths, including from SNRs and HNRs.  In addition, CRs may also be injected from 
disk-driven outflows and weak jets from AGNs \citep{murase2014diffuse,tamborra2014star,Liu:2017bjr}. 
The CR contributions from these sources, which would be additional to CRs from the mergers
considered here, are significantly model-dependent, and we do not attempt here a quantification
of their relative importance.}
 
One important factor that may influence the final results is the CR power-law index $s$, since the 
factor $\varepsilon^{2-s}$, the maximal CR energy as well as a new $\mathcal C=
(({\varepsilon_p^{\rm max})}^{2-s}-{(\varepsilon_p^{\rm min})}^{2-s})/(2-s)$ 
are required to correct the Eq. \ref{eq:crinput} when $s$ deviates from $2.0$. 
As presented in Eq. \ref{eq:crinput}, we assume that the shock is non-radiative and {\ccminor infinitesimally thin} 
and hence the Fermi first order acceleration in the strong shock limit implies $s=2$. 
However, a finite width of the shock can steepen the spectrum to $s\gtrsim2.0$, while a radiative 
shock would produce a CR spectrum with a power-law index lower than $2.0$.  
For radiative shocks, {\ccminor\cite{blandford1987particle} showed} that the power-law index of the accelerated CRs is 
$s=(\phi+2)/(\phi-1)$ where $\phi$ is the compression ratio. Moreover, \cite{yamazaki2006tev} assume 
$\phi=7$ and $s=1.5$ as fiducial values when studying the radiative cooling of SNR shocks. Note that 
the compression ratio for radiative shocks with an isothermal adiabatic index $\gamma=1$ can be 
written as $\phi=M^2\gg 1$, where $M$ is the upstream Mach number. We asume thus $s=1$ in this 
extreme case. To illustrate how the neutrino spectra are affected by the radiative cooling and/or 
the width of the shocks, we plot in Fig. \ref{fig:CRindex} the neutrino fluxes for four cases 
$s=2.2 ,2.0,\ 1.5$ and $1.03$ which correspond to $\phi=3.5,\ 4,\ 7,\ 100$, respectively.
\begin{figure}
\includegraphics[width=0.45\textwidth]{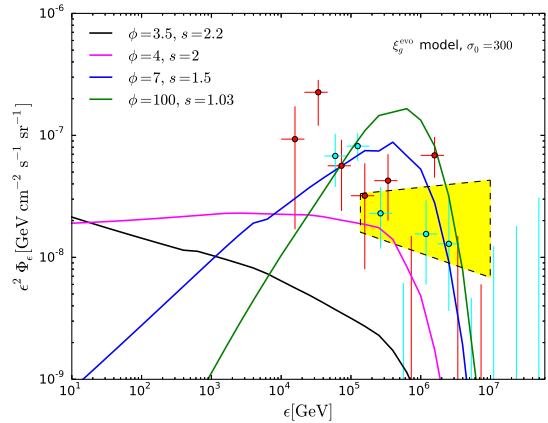}
\caption{The neutrino fluxes for different compression ratios and CR power-law indices. The black, magenta, 
blue and greens lines correspond to the power-law indices $s=2.2,\ 2.0,\ 1.5$ and $1.03$.}
\label{fig:CRindex}
\end{figure}
As can be seen, {\ccminor a harder CR power-law index} (lower $s$) will produce more high-energy neutrinos. 
Thus, in principle, a 
mildly radiative-cooling shock ($1.5\leq s\leq2$) can more easily achieve the high-energy neutrino flux in the range 
$10\ {\rm TeV}$ to $\sim\rm PeV$. On the other hand, $s\gtrsim2.1-2.2$ is disfavored because of the damping factor 
$\varepsilon^{2-s}$, which is consistent with previous work~\citep{murase2013testing}. 
Note that the hard spectrum is expected for the cold gas environment that would be valid in sufficiently 
low-mass halo mergers.  If the temperature is so high, the Mach number is expected to be low, as expected 
for cluster mergers. In this case, the spectrum is softer for massive clusters, and details are beyond 
the scope of this work. 

Additional contributions may arise from the galaxies moving through the cluster  or dark matter halo,
as their hypersonic peculiar motion will result in a shock as they plow through the intra-cluster gas, 
which as a result can also contribute to the diffuse $\gamma$ rays and neutrinos. Supposing as an extreme 
case that the loss of the galaxies' kinetic energy due to the gravitational drag is completely converted 
into CR energy, we estimate a CR energy budget of
\begin{linenomath*}
\begin{equation}
\varepsilon_pQ_{\varepsilon_p}^{\rm (IGM)}=\epsilon_p \mathcal C^{-1}_{\rm IGM}\int dM_h\frac{4\pi G^2M_h^2\rho_{cl}}{v_s}\frac{dN}{dM_h}\ln\left(\frac{R_{\rm cl}}{R_{g}}\right),
\label{eq:IGMbudget}
\end{equation} 
\end{linenomath*}
which is three orders of magnitudes lower than halo mergers estimated in the previous sections, 
because of the tenuous intergalactic gas density. Hence, these shocks contribute only {\ccminor a relatively 
small amount of} diffuse neutrinos and are negligible compared to the mergers.  

\section{Summary}
In summary, we found that the CR luminosity density by halo mergers can be comparable to that from starburst galaxies,
{\ccmajor which can be expected from galaxy mergers.}
In particular, the CR input from galaxy mergers and cluster/group mergers is comparable in the local universe, and 
the former is more important at higher redshifts, $z\gtrsim2$ (see the dashed and dash-dotted lines in 
Figure~(\ref{fig:CRinput}). {\ccminor This emphasizes the importance of our results for CR reservoir models. 
We have considered the neutrino and $\gamma$-ray production in galaxy-galaxy and cluster/group merger environments 
and found that such mergers could explain a large portion of the IceCube diffuse neutrino flux.} 
Since many more galaxy-scale, low-mass halo mergers occur at relatively high redshifts, the contribution to 
the diffuse $\gamma$-ray background observed by $Fermi$ is more suppressed, due to the $\gamma\gamma$ absorption. 
Despite the various uncertainties due to the lack of high redshift observations of the galactic and cluster 
morphologies, the gas distribution and the galactic/intergalactic magnetic fields, some of the crucial and 
sensitive parameters including the gas fraction $\xi_g^{\rm evo}$ are relatively well constrained. 
The parameter space left for variance of both the neutrino and $\gamma$-ray spectra is restricted by our results, 
as demonstrated in Figs. \ref{fig:const_gas_fraction} and \ref{fig:evo_gas_fraction}. 
One of the large uncertainties comes from the spectral index, and we demonstrated the cases of harder 
CR spectral indices, $1.5\lesssim s\lesssim2$, which could be expected in strong radiative shocks. 

One of the predictions of the halo merger model is that the effective number density of these sources is 
expected to be $\sim10^{-5}~{\rm Mpc}^{-3}$, which is similar to the number density of starburst galaxies 
and AGN with disk-driven outflows. {\km The present model is testable in the sense that such halo merger sources are detectable} 
with next-generation detectors such as IceCube-Gen2 via searches for multiplets, auto-correlation, and 
cross-correlation signals \citep{Murase:2016gly}. {\km One must keep in mind that the contributions from
galaxy/halo mergers are degenerate with those from other possibilities, such as the starburst and AGN contributions, since a 
large fraction of starburst and AGN activities can be induced by these mergers. To distinguish among these models, 
cross-correlation or auto-correlation studies in neutrinos and $\gamma$-rays should be useful. 
Also, to identify the merging sources, it will be important to investigate these sources at multi-wavelengths.}
% {\ccmajor Our on-going investigation towards the EM signature generated from the post-shock bremsstrahlung 
%radiation and the secondary electrons can provide guidance for future EM observations. }
%

\section{Acknowledgements} 
We are grateful to Zhao-Wei Zhang and Kazumi Kashiyama for useful discussions. This research was partially supported by 
NASA NNX13AH50G (C-C.Y., P.M.), the Alfred P. Sloan Foundation and NSF grant No. PHY-1620777 (K.M.)
and NSF grant No. AST-1517363 (D.J.).

\section*{Appendix A: the merger rate} 
\begin{figure}\centering
{\includegraphics[width=0.45\textwidth]{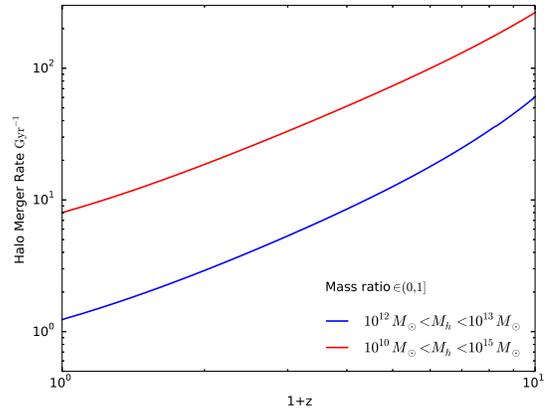}}
\caption{Merger rates from Eq. (22). The blue line represents the whole mass range ($10^{10}~{\rm M}_\odot\sim10^{15}~\rm M_\odot$) and the red line corresponds to $10^{12}~{\rm M}_\odot<M_h<10^{13}~\rm M_\odot$}.
\label{fig:mergrate}
\end{figure}

In this section we present a comparison between our halo merger rate with the Illustris simulations \citep{rodriguez2015merger}. In our calculation, we assign a mean merger probability $P(M,z)=\exp(-t_{\rm merger}/t_{\rm age})$  to each dark matter halo during $t_{\rm age}$. Here, $t_{\rm merger}$, which can be obtained from the second equation of Eq.(\ref{eq:times3}), averages all possible mass ratios, e.g. $\zeta\in(0,1]$. In our calculation, we do not need to use the cumulative merger rate over mass directly, instead the factor $\frac{dN}{dM}\frac{P(M,z)}{t_{\rm age}}$ in the integrand of Eq. (\ref{eq:crinput}) is used to illustrate the number of mergers for a halo with mass $M$ and at redshift $z$. However, in order to compare our results with the simulations, it is worthwhile to estimate the average cumulative merger rate using our approach, 
\begin{linenomath*}
\begin{equation}
\mathcal R(z)=\frac{\int \frac{dN}{dM}\frac{P(M,z)}{t_{age}}dM}{\int \frac{dN}{dM}dM}.
\end{equation} 
\end{linenomath*}
The merger rate is shown in Fig. \ref{fig:mergrate} where the blue line represents the whole mass range ($10^{10}~{\rm M}_\odot\sim10^{15}~{\rm M}_\odot$) and the red line corresponds to $10^{12}~{\rm M}_\odot<M_h<10^{13}~{\rm M}_\odot$. In both cases, the mass ratio covers the entire interval as in the middle equation of Eq. (\ref{eq:times3}) 
{{\ccminor which is integrated over}} $\zeta$ from $0$ to $1$. The merger rate given by Illustris simulations is shown in the lower panel.

One can compare our results with solid black lines in the right panel of Fig. 2 in \citet{rodriguez2015merger}, since the increase in the merger rates given by simulations (as shown as colored lines) seen at low redshifts is due to a limitation of the splitting algorithm. As can be seen, our merger rate in the same mass interval is comparable to the merger rate in the right panel with the mass ratio $\geq1/1000$. Considering that we are using a totally different method and this approach is primarily designed to evaluate the merger probabilities of halos of various masses and at different redshifts, the moderate degree of discrepancy can be considered acceptable.

\bibliography{ref.bib}

\begin{thebibliography}{}
\expandafter\ifx\csname natexlab\endcsname\relax\def\natexlab#1{#1}\fi
\providecommand{\url}[1]{\href{#1}{#1}}
\providecommand{\dodoi}[1]{doi:~\href{http://doi.org/#1}{\nolinkurl{#1}}}
\providecommand{\doeprint}[1]{\href{http://ascl.net/#1}{\nolinkurl{http://ascl.net/#1}}}
\providecommand{\doarXiv}[1]{\href{https://arxiv.org/abs/#1}{\nolinkurl{https://arxiv.org/abs/#1}}}

\bibitem[{Aartsen {et~al.}(2013{\natexlab{a}})}]{Aartsen:2013bka}
Aartsen, M., {et~al.} 2013{\natexlab{a}}, Phys.Rev.Lett., 111, 021103

\bibitem[{Aartsen {et~al.}(2013{\natexlab{b}})}]{Aartsen:2013jdh}
---. 2013{\natexlab{b}}, Science, 342, 1242856

\bibitem[{Aartsen {et~al.}(2014)}]{Aartsen:2014gkd}
---. 2014, Phys.Rev.Lett., 113, 101101

\bibitem[{Aartsen {et~al.}(2015)Aartsen, Abraham, Ackermann, Adams, Aguilar,
  Ahlers, Ahrens, Altmann, Anderson, Archinger, {et~al.}}]{aartsen2015combined}
Aartsen, M., Abraham, K., Ackermann, M., {et~al.} 2015, The Astrophysical
  Journal, 809, 98

\bibitem[{Aartsen {et~al.}(2016)Aartsen, Abraham, Ackermann, Adams, Aguilar,
  Ahlers, Ahrens, Altmann, Andeen, Anderson, {et~al.}}]{aartsen2016observation}
---. 2016, The Astrophysical Journal, 833, 3

\bibitem[{Aartsen {et~al.}(2017)}]{Aartsen:2017mau}
Aartsen, M.~G., {et~al.} 2017, arXiv: 1710.01191

\bibitem[{Abeysekara {et~al.}(2017{\natexlab{a}})}]{Abeysekara:2017wzt}
Abeysekara, A.~U., {et~al.} 2017{\natexlab{a}}, Astrophys. J., 842, 85

\bibitem[{Abeysekara {et~al.}(2017{\natexlab{b}})}]{Abeysekara:2017jxs}
---. 2017{\natexlab{b}}, arXiv: 1710.10288

\bibitem[{{Ackermann} {et~al.}(2016){Ackermann}, {Ajello}, {Albert}, {Atwood},
  {Baldini}, \& et~al.}]{Fermi16}
{Ackermann}, M., {Ajello}, M., {Albert}, A., {et~al.} 2016, Physical Review
  Letters, 116, 151105

\bibitem[{Ackermann {et~al.}(2015)Ackermann, Ajello, Albert, Atwood, Baldini,
  Ballet, Barbiellini, Bastieri, Bechtol, Bellazzini,
  {et~al.}}]{ackermann2015spectrum}
Ackermann, M., Ajello, M., Albert, A., {et~al.} 2015, The Astrophysical
  Journal, 799, 86

\bibitem[{Ahlers \& Murase(2014)}]{Ahlers:2013xia}
Ahlers, M., \& Murase, K. 2014, Phys. Rev., D90, 023010

\bibitem[{Alvarez-Mu{\~n}iz \& M{\'e}sz{\'a}ros(2004)}]{alvarez2004high}
Alvarez-Mu{\~n}iz, J., \& M{\'e}sz{\'a}ros, P. 2004, Physical Review D, 70,
  123001

\bibitem[{Anchordoqui {et~al.}(2008)Anchordoqui, Hooper, Sarkar, \&
  Taylor}]{anchordoqui2008high}
Anchordoqui, L.~A., Hooper, D., Sarkar, S., \& Taylor, A.~M. 2008,
  Astroparticle Physics, 29, 1

\bibitem[{Anchordoqui {et~al.}(2014)Anchordoqui, Paul, da~Silva, Torres, \&
  Vlcek}]{anchordoqui2014icecube}
Anchordoqui, L.~A., Paul, T.~C., da~Silva, L.~H., Torres, D.~F., \& Vlcek,
  B.~J. 2014, Physical Review D, 89, 127304

\bibitem[{Apel {et~al.}(2017)}]{Apel:2017ocm}
Apel, W.~D., {et~al.} 2017, Astrophys. J., 848, 1

\bibitem[{Baerwald {et~al.}(2013)Baerwald, Bustamante, \&
  Winter}]{baerwald2013uhecr}
Baerwald, P., Bustamante, M., \& Winter, W. 2013, The Astrophysical Journal,
  768, 186

\bibitem[{Bahcall \& Soneira(1980)}]{bahcall1980universe}
Bahcall, J.~N., \& Soneira, R.~M. 1980, The Astrophysical Journal Supplement
  Series, 44, 73

\bibitem[{Barnaby \& Thronson~Jr(1992)}]{barnaby1992distribution}
Barnaby, D., \& Thronson~Jr, H.~A. 1992, The Astronomical Journal, 103, 41

\bibitem[{Bechtol {et~al.}(2017)Bechtol, Ahlers, Di~Mauro, Ajello, \&
  Vandenbroucke}]{bechtol2017evidence}
Bechtol, K., Ahlers, M., Di~Mauro, M., Ajello, M., \& Vandenbroucke, J. 2017,
  The Astrophysical Journal, 836, 47

\bibitem[{Becker~Tjus {et~al.}(2014)Becker~Tjus, Eichmann, Halzen, Kheirandish,
  \& Saba}]{Tjus:2014dna}
Becker~Tjus, J., Eichmann, B., Halzen, F., Kheirandish, A., \& Saba, S.~M.
  2014, Phys. Rev., D89, 123005

\bibitem[{Behroozi {et~al.}(2013)Behroozi, Wechsler, \&
  Conroy}]{behroozi2013average}
Behroozi, P.~S., Wechsler, R.~H., \& Conroy, C. 2013, The Astrophysical
  Journal, 770, 57

\bibitem[{Berezinsky \& Smirnov(1975)}]{berezinsky1975cosmic}
Berezinsky, V., \& Smirnov, A.~Y. 1975, Astrophysics and Space Science, 32, 461

\bibitem[{Berrington \& Dermer(2003)}]{Berrington:2002bh}
Berrington, R.~C., \& Dermer, C.~D. 2003, Astrophys. J., 594, 709

\bibitem[{Blanco \& Hooper(2017)}]{Blanco:2017bgl}
Blanco, C., \& Hooper, D. 2017, JCAP, 1712, 017

\bibitem[{Blandford \& Eichler(1987)}]{blandford1987particle}
Blandford, R., \& Eichler, D. 1987, Physics Reports, 154, 1

\bibitem[{Bonvin {et~al.}(2017)Bonvin, Courbin, Suyu, Marshall, Rusu, Sluse,
  Tewes, Wong, Collett, Fassnacht, {et~al.}}]{bonvin2017h0licow}
Bonvin, V., Courbin, F., Suyu, S., {et~al.} 2017, Monthly Notices of the Royal
  Astronomical Society, 465, 4914

\bibitem[{Bustamante {et~al.}(2014)Bustamante, Baerwald, Murase, \&
  Winter}]{Bustamante:2014oka}
Bustamante, M., Baerwald, P., Murase, K., \& Winter, W. 2014

\bibitem[{Chakraborty \& Izaguirre(2015)}]{Chakraborty:2015sta}
Chakraborty, S., \& Izaguirre, I. 2015, Phys. Lett., B745, 35

\bibitem[{Chang {et~al.}(2015)Chang, Liu, \& Wang}]{chang2015star}
Chang, X.-C., Liu, R.-Y., \& Wang, X.-Y. 2015, The Astrophysical Journal, 805,
  95

\bibitem[{{Chang} {et~al.}(2016){Chang}, {Liu}, \& {Wang}}]{Chang+16grbnugam}
{Chang}, X.-C., {Liu}, R.-Y., \& {Wang}, X.-Y. 2016, \apj, 825, 148

\bibitem[{Chang \& Wang(2014)}]{chang2014diffuse}
Chang, X.-C., \& Wang, X.-Y. 2014, The Astrophysical Journal, 793, 131

\bibitem[{Charbonnel {et~al.}(2011)Charbonnel, Montmerle, \&
  Bournaud}]{charbonnel2011star}
Charbonnel, C., Montmerle, T., \& Bournaud, F. 2011, European Astronomical
  Society Publications Series, 51, 107

\bibitem[{Clowe {et~al.}(2004)Clowe, Gonzalez, \& Markevitch}]{clowe2004weak}
Clowe, D., Gonzalez, A., \& Markevitch, M. 2004, The Astrophysical Journal,
  604, 596

\bibitem[{Crocker(2012)}]{ref8}
Crocker, R.~M. 2012, Monthly Notices of the Royal Astronomical Society, 423,
  3512

\bibitem[{Croston {et~al.}(2008)Croston, Pratt, B{\"o}hringer, Arnaud,
  Pointecouteau, Ponman, Sanderson, Temple, Bower, \&
  Donahue}]{croston2008galaxy}
Croston, J., Pratt, G., B{\"o}hringer, H., {et~al.} 2008, Astronomy \&
  Astrophysics, 487, 431

\bibitem[{Davis \& Peebles(1977)}]{davis1977integration}
Davis, M., \& Peebles, P. 1977, The Astrophysical Journal Supplement Series,
  34, 425

\bibitem[{Davis \& Peebles(1983)}]{davis1983survey}
---. 1983, The Astrophysical Journal, 267, 465

\bibitem[{Denton \& Tamborra(2017)}]{Denton:2017jwk}
Denton, P.~B., \& Tamborra, I. 2017, arXiv: 1711.00470

\bibitem[{Dermer {et~al.}(2014)Dermer, Murase, \& Inoue}]{dermer2014photopion}
Dermer, C.~D., Murase, K., \& Inoue, Y. 2014, Journal of High Energy
  Astrophysics, 3, 29

\bibitem[{Drury(1983)}]{drury1983introduction}
Drury, L.~O. 1983, Reports on Progress in Physics, 46, 973

\bibitem[{Eke {et~al.}(1998)Eke, Navarro, \& Frenk}]{ref9}
Eke, V.~R., Navarro, J.~F., \& Frenk, C.~S. 1998, The Astrophysical Journal,
  503, 569

\bibitem[{Fakhouri {et~al.}(2010)Fakhouri, Ma, \& Boylan-Kolchin}]{ref11}
Fakhouri, O., Ma, C.-P., \& Boylan-Kolchin, M. 2010, Monthly Notices of the
  Royal Astronomical Society, 406, 2267

\bibitem[{Fang \& Murase(2018)}]{fang2018linking}
Fang, K., \& Murase, K. 2018, Nature Physics, 1

\bibitem[{Fialkov \& Loeb(2016)}]{ref10}
Fialkov, A., \& Loeb, A. 2016, arXiv preprint arXiv:1611.01386

\bibitem[{Finke {et~al.}(2010)Finke, Razzaque, \& Dermer}]{finke2010modeling}
Finke, J.~D., Razzaque, S., \& Dermer, C.~D. 2010, The Astrophysical Journal,
  712, 238

\bibitem[{Fujita \& Sarazin(2001)}]{fujita2001nonthermal}
Fujita, Y., \& Sarazin, C.~L. 2001, The Astrophysical Journal, 563, 660

\bibitem[{Gaisser \& Halzen(2014)}]{doi:10.1146/annurev-nucl-102313-025321}
Gaisser, T., \& Halzen, F. 2014, Annual Review of Nuclear and Particle Science,
  64, 101

\bibitem[{Groth \& Peebles(1977)}]{groth1977statistical}
Groth, E.~J., \& Peebles, P. 1977, The Astrophysical Journal, 217, 385

\bibitem[{Gupta \& Zhang(2007)}]{gupta2007neutrino}
Gupta, N., \& Zhang, B. 2007, Astroparticle Physics, 27, 386

\bibitem[{Halzen(2016)}]{Halzen:2016gng}
Halzen, F. 2016, Nature Phys., 13, 232

\bibitem[{Halzen \& Zas(1997)}]{Halzen:1997hw}
Halzen, F., \& Zas, E. 1997, Astrophys. J., 488, 669

\bibitem[{Hamilton {et~al.}(1991)Hamilton, Kumar, Lu, \&
  Matthews}]{hamilton1991reconstructing}
Hamilton, A., Kumar, P., Lu, E., \& Matthews, A. 1991, The Astrophysical
  Journal, 374, L1

\bibitem[{Hawkins {et~al.}(2003)Hawkins, Maddox, Cole, Lahav, Madgwick,
  Norberg, Peacock, Baldry, Baugh, Bland-Hawthorn, {et~al.}}]{hawkins20032df}
Hawkins, E., Maddox, S., Cole, S., {et~al.} 2003, Monthly Notices of the Royal
  Astronomical Society, 346, 78

\bibitem[{Inoue {et~al.}(2013)Inoue, Inoue, Kobayashi, Makiya, Niino, \&
  Totani}]{inoue2013extragalactic}
Inoue, Y., Inoue, S., Kobayashi, M.~A., {et~al.} 2013, The Astrophysical
  Journal, 768, 197

\bibitem[{Jenkins {et~al.}(2001)Jenkins, Frenk, White, Colberg, Cole, Evrard,
  Couchman, \& Yoshida}]{ref6}
Jenkins, A., Frenk, C., White, S.~D., {et~al.} 2001, Monthly Notices of the
  Royal Astronomical Society, 321, 372

\bibitem[{Jing {et~al.}(1998)Jing, Mo, \& B{\"o}rner}]{jing1998spatial}
Jing, Y., Mo, H., \& B{\"o}rner, G. 1998, The Astrophysical Journal, 494, 1

\bibitem[{Kafexhiu {et~al.}(2014)Kafexhiu, Aharonian, Taylor, \&
  Vila}]{kafexhiu2014parametrization}
Kafexhiu, E., Aharonian, F., Taylor, A.~M., \& Vila, G.~S. 2014, Physical
  Review D, 90, 123014

\bibitem[{Kashiyama \& M{\'e}sz{\'a}ros(2014)}]{kashiyama2014galaxy}
Kashiyama, K., \& M{\'e}sz{\'a}ros, P. 2014, The Astrophysical Journal Letters,
  790, L14

\bibitem[{Keeney {et~al.}(2006)Keeney, Danforth, Stocke, Penton, Shull, \&
  Sembach}]{ref7}
Keeney, B.~A., Danforth, C.~W., Stocke, J.~T., {et~al.} 2006, The Astrophysical
  Journal, 646, 951

\bibitem[{Kimura {et~al.}(2015)Kimura, Murase, \& Toma}]{Kimura:2014jba}
Kimura, S.~S., Murase, K., \& Toma, K. 2015, Astrophys. J., 806, 159

\bibitem[{Kylafis \& Bahcall(1987)}]{kylafis1987dust}
Kylafis, N.~D., \& Bahcall, J.~N. 1987, The Astrophysical Journal, 317, 637

\bibitem[{Lamastra {et~al.}(2017)Lamastra, Menci, Fiore, Antonelli,
  Colafrancesco, Guetta, \& Stamerra}]{Lamastra:2017iyo}
Lamastra, A., Menci, N., Fiore, F., {et~al.} 2017, Astron. Astrophys., 607, A18

\bibitem[{Lewis {et~al.}(2000)Lewis, Challinor, \& Lasenby}]{camb}
Lewis, A., Challinor, A., \& Lasenby, A. 2000, Astrophys. J., 538, 473

\bibitem[{Lisanti {et~al.}(2016)Lisanti, Mishra-Sharma, Necib, \&
  Safdi}]{Lisanti:2016jub}
Lisanti, M., Mishra-Sharma, S., Necib, L., \& Safdi, B.~R. 2016, Astrophys. J.,
  832, 117

\bibitem[{Liu {et~al.}(2017)Liu, Murase, Inoue, Ge, \& Wang}]{Liu:2017bjr}
Liu, R.-Y., Murase, K., Inoue, S., Ge, C., \& Wang, X.-Y. 2017,
  arXiv:1712.10168

\bibitem[{Loeb \& Waxman(2006)}]{loeb2006cumulative}
Loeb, A., \& Waxman, E. 2006, Journal of Cosmology and Astroparticle Physics,
  2006, 003

\bibitem[{Mannheim(1995)}]{Mannheim:1995mm}
Mannheim, K. 1995, Astropart. Phys., 3, 295

\bibitem[{Markevitch {et~al.}(2004)Markevitch, Gonzalez, Clowe, Vikhlinin,
  Forman, Jones, Murray, \& Tucker}]{markevitch2004direct}
Markevitch, M., Gonzalez, A., Clowe, D., {et~al.} 2004, The Astrophysical
  Journal, 606, 819

\bibitem[{Meszaros \& Waxman(2001)}]{meszaros2001tev}
Meszaros, P., \& Waxman, E. 2001, Physical Review Letters, 87, 171102

\bibitem[{Murase(2008)}]{murase2008prompt}
Murase, K. 2008, Physical Review D, 78, 101302

\bibitem[{Murase {et~al.}(2013)Murase, Ahlers, \& Lacki}]{murase2013testing}
Murase, K., Ahlers, M., \& Lacki, B.~C. 2013, Physical Review D, 88, 121301

\bibitem[{Murase \& Beacom(2012)}]{murase2012constraining}
Murase, K., \& Beacom, J.~F. 2012, Journal of Cosmology and Astroparticle
  Physics, 2012, 043

\bibitem[{Murase {et~al.}(2016)Murase, Guetta, \& Ahlers}]{Murase:2015xka}
Murase, K., Guetta, D., \& Ahlers, M. 2016, Phys. Rev. Lett., 116, 071101

\bibitem[{Murase {et~al.}(2008)Murase, Inoue, \& Nagataki}]{murase2008cosmic}
Murase, K., Inoue, S., \& Nagataki, S. 2008, The Astrophysical Journal Letters,
  689, L105

\bibitem[{Murase {et~al.}(2014)Murase, Inoue, \& Dermer}]{murase2014diffuse}
Murase, K., Inoue, Y., \& Dermer, C.~D. 2014, Physical Review D, 90, 023007

\bibitem[{Murase \& Ioka(2013)}]{murase2013tev}
Murase, K., \& Ioka, K. 2013, Physical Review Letters, 111, 121102

\bibitem[{Murase {et~al.}(2006)Murase, Ioka, Nagataki, \&
  Nakamura}]{murase2006high}
Murase, K., Ioka, K., Nagataki, S., \& Nakamura, T. 2006, The Astrophysical
  Journal Letters, 651, L5

\bibitem[{Murase \& Waxman(2016)}]{Murase:2016gly}
Murase, K., \& Waxman, E. 2016, Phys. Rev., D94, 103006

\bibitem[{Murray {et~al.}(2013)Murray, Power, \& Robotham}]{ref5}
Murray, S., Power, C., \& Robotham, A. 2013, Monthly Notices of the Royal
  Astronomical Society: Letters, 434, L61

\bibitem[{Padovani {et~al.}(2015)Padovani, Petropoulou, Giommi, \&
  Resconi}]{Padovani:2015mba}
Padovani, P., Petropoulou, M., Giommi, P., \& Resconi, E. 2015, Mon. Not. Roy.
  Astron. Soc., 452, 1877

\bibitem[{Peacock \& Dodds(1996)}]{peacock1996non}
Peacock, J., \& Dodds, S. 1996, Monthly Notices of the Royal Astronomical
  Society, 280, L19

\bibitem[{Peebles(1974)}]{peebles1974gravitational}
Peebles, P.~J. 1974, The Astrophysical Journal, 189, L51

\bibitem[{Peebles(1980)}]{peebles1980large}
Peebles, P. J.~E. 1980, The large-scale structure of the universe (Princeton
  university press)

\bibitem[{Petropoulou {et~al.}(2015)Petropoulou, Dimitrakoudis, Padovani,
  Mastichiadis, \& Resconi}]{petropoulou2015photohadronic}
Petropoulou, M., Dimitrakoudis, S., Padovani, P., Mastichiadis, A., \& Resconi,
  E. 2015, Monthly Notices of the Royal Astronomical Society, 448, 2412

\bibitem[{Press \& Schechter(1974)}]{press1974formation}
Press, W.~H., \& Schechter, P. 1974, The Astrophysical Journal, 187, 425

\bibitem[{Reed {et~al.}(2006)Reed, Bower, Frenk, Jenkins, \& Theuns}]{ref4}
Reed, D.~S., Bower, R., Frenk, C.~S., Jenkins, A., \& Theuns, T. 2006, Monthly
  Notices of the Royal Astronomical Society, 374, 2

\bibitem[{Rodriguez-Gomez {et~al.}(2015)Rodriguez-Gomez, Genel, Vogelsberger,
  Sijacki, Pillepich, Sales, Torrey, Snyder, Nelson, Springel,
  {et~al.}}]{rodriguez2015merger}
Rodriguez-Gomez, V., Genel, S., Vogelsberger, M., {et~al.} 2015, Monthly
  Notices of the Royal Astronomical Society, 449, 49

\bibitem[{Sargent {et~al.}(2014)Sargent, Daddi, B{\'e}thermin, Aussel, Magdis,
  Hwang, Juneau, Elbaz, \& Da~Cunha}]{sargent2014regularity}
Sargent, M.~T., Daddi, E., B{\'e}thermin, M., {et~al.} 2014, The Astrophysical
  Journal, 793, 19

\bibitem[{Senno {et~al.}(2015)Senno, M{\'e}sz{\'a}ros, Murase, Baerwald, \&
  Rees}]{senno2015extragalactic}
Senno, N., M{\'e}sz{\'a}ros, P., Murase, K., Baerwald, P., \& Rees, M.~J. 2015,
  The Astrophysical Journal, 806, 24

\bibitem[{Senno {et~al.}(2016)Senno, Murase, \&
  M{\'e}sz{\'a}ros}]{senno2016choked}
Senno, N., Murase, K., \& M{\'e}sz{\'a}ros, P. 2016, Physical Review D, 93,
  083003

\bibitem[{Sheth \& Tormen(1999)}]{ref3}
Sheth, R.~K., \& Tormen, G. 1999, Monthly Notices of the Royal Astronomical
  Society, 308, 119

\bibitem[{Shibuya {et~al.}(2015)Shibuya, Ouchi, \&
  Harikane}]{shibuya2015morphologies}
Shibuya, T., Ouchi, M., \& Harikane, Y. 2015, The Astrophysical Journal
  Supplement Series, 219, 15

\bibitem[{{Smith} {et~al.}(2003){Smith}, {Peacock}, {Jenkins}, {White},
  {Frenk}, {Pearce}, {Thomas}, {Efstathiou}, \& {Couchman}}]{smith/etal:2003}
{Smith}, R.~E., {Peacock}, J.~A., {Jenkins}, A., {et~al.} 2003, Monthly Notices
  of the Royal Astronomical Society, 341, 1311

\bibitem[{Stecker(2013)}]{stecker2013pev}
Stecker, F.~W. 2013, Physical Review D, 88, 047301

\bibitem[{Stecker {et~al.}(1991)Stecker, Done, Salamon, \&
  Sommers}]{stecker1991high}
Stecker, F.~W., Done, C., Salamon, M.~H., \& Sommers, P. 1991, Physical Review
  Letters, 66, 2697

\bibitem[{Tamborra \& Ando(2016)}]{tamborra2016inspecting}
Tamborra, I., \& Ando, S. 2016, Physical Review D, 93, 053010

\bibitem[{Tamborra {et~al.}(2014)Tamborra, Ando, \& Murase}]{tamborra2014star}
Tamborra, I., Ando, S., \& Murase, K. 2014, Journal of Cosmology and
  Astroparticle Physics, 2014, 043

\bibitem[{Van Der~Kruit \& Searle(1981)}]{van1981surface}
Van Der~Kruit, P., \& Searle, L. 1981, Astronomy and Astrophysics, 95, 116

\bibitem[{Wang \& Dai(2009)}]{wang2009prompt}
Wang, X.-Y., \& Dai, Z.-G. 2009, The Astrophysical Journal Letters, 691, L67

\bibitem[{Waxman \& Bahcall(1997)}]{waxman1997high}
Waxman, E., \& Bahcall, J. 1997, Physical Review Letters, 78, 2292

\bibitem[{Xiao \& Dai(2014)}]{xiao2014neutrino}
Xiao, D., \& Dai, Z. 2014, The Astrophysical Journal, 790, 59

\bibitem[{Xiao \& Dai(2015)}]{xiao2015tev}
---. 2015, The Astrophysical Journal, 805, 137

\bibitem[{{Xiao} {et~al.}(2016){Xiao}, {M{\'e}sz{\'a}ros}, {Murase}, \&
  {Dai}}]{Xiao+16nuhn}
{Xiao}, D., {M{\'e}sz{\'a}ros}, P., {Murase}, K., \& {Dai}, Z.-G. 2016, \apj,
  826, 133

\bibitem[{Yamazaki {et~al.}(2006)Yamazaki, Kohri, Bamba, Yoshida, Tsuribe, \&
  Takahara}]{yamazaki2006tev}
Yamazaki, R., Kohri, K., Bamba, A., {et~al.} 2006, Monthly Notices of the Royal
  Astronomical Society, 371, 1975

\bibitem[{Zehavi {et~al.}(2002)Zehavi, Blanton, Frieman, Weinberg, Mo, Strauss,
  Anderson, Annis, Bahcall, Bernardi, {et~al.}}]{zehavi2002galaxy}
Zehavi, I., Blanton, M.~R., Frieman, J.~A., {et~al.} 2002, The Astrophysical
  Journal, 571, 172

\end{thebibliography}
\end{document}